\documentclass[%
 preprint,onecolumn,
 amsmath,amssymb,
 aps,
]{revtex4-1}
\usepackage{graphicx,color}
\usepackage{here}
\usepackage{caption}
\usepackage[skip=0cm]{subcaption}
\usepackage{dcolumn}
\usepackage{bm}

\begin{document}


\title{Constraints for the thawing and freezing potentials}
\author{Tetsuya Hara}
\email{hara@cc.kyoto-su.ac.jp}
\affiliation{Department of Physics, Kyoto Sangyo University, Kyoto 603-8555, Japan}
\author{Anna Suzuki}%
\affiliation{Department of Physics, Kyoto Sangyo University, Kyoto 603-8555, Japan}
\author{Shogo Saka}%
\affiliation{Department of Physics, Kyoto Sangyo University, Kyoto 603-8555, Japan}
\author{Takuma Tanigawa} 
\affiliation{Department of Physics, Kyoto Sangyo University, Kyoto 603-8555, Japan}


\begin{abstract}
\hspace{5mm} We study the accelerating present universe in terms of the time evolution of the equation of state $w(z)$ (redshift $z$) 
due to thawing and freezing scalar potentials in the quintessence model.  The values of $dw/da$ and $d^2w/da^2$ at scale factor of $a = 1$ are associated 
with two parameters of each potential.  For five type scalar potentials, the scalar fields $Q$ and $w$ as function of time $t$ and/or $z$ 
are numerically calculated under the fixed boundary condition of $w(z=0)=-1+\Delta$.  
The observational constraint $w_{obs}$ (Planck 2015) is imposed to test whether the numerical $w(z)$ is in $w_{obs}$.  
Some solutions show the thawing features in the freezing potentials.  Mutually exclusive allowed regions in $dw/da$ vs. $d^2w/da^2$ diagram are obtained 
in order to identify the likely scalar potential and even the potential parameters by future observational tests. 
\end{abstract}	

\maketitle

\section{Introduction}
\hspace{1mm}Although almost two decades have passed since the observation of the accelerating universe, dark energy is still left as one of 
the most mysterious problems in physics \citep{1}. We investigate the scalar field in quintessence scenario how relevant 
it is to the dark energy \citep{2,3,4}. 
 In this scenario, if the potential of the scalar field has $n$ independent 
  parameters, the first, second,  $\cdots $ and $n$-th time derivatives observation of the equation of state $w$ are enough to specify the scalar potentials
  and to predict the higher derivatives 
  \citep{5,6}.  
It has been said that the potentials could be classified to freezing type and thawing type \citep{7}.  
The equation of state $w$ will approach to $-1$ in the freezing type and $w$ will depart from $-1$ in the thawing model.  
We have adopted three types for freezing potential as $ V=M^{4+\alpha}/Q^{\alpha}$ (inverse power law; $\alpha > 0$) \citep{8}, 
 $ V=M^4\exp(\beta M/Q)$ (exponential; $\beta > 0$), and $V=M^{4+\gamma}/Q^{\gamma}\exp (4\pi Q^2)$ (mixed; $\gamma > 0$) \citep{9}, 
 and two types for thawing potential as  $V=M^4(\cos (Q/f)+1))$ (cosine)  and $V=M^4\exp(-Q^2/\sigma^2)$ (Gaussian) \citep{10}.  
 They are rather simple potentials and have two parameters to specify the form of potential, respectively.

By adopting the values of  
the derivatives $dw/da$ and $d^2w/da^2$ at the scale factor of $a = 1$, the characteristic form parameters of each potential could be fixed \citep{5,6}
and then the theoretical time development of the scalar field $Q$ in the potential could be numerically calculated from the present
to the past (and then the reverse).

Through the investigation of the adopted potentials, we have found that there is a forbidden region in the first and second derivatives of $w$ for each potential.  
  Although there is a lot of uncertainty at the moment, it is estimated the allowed regions for potentials in the derivative space of $w$ such as $dw/da-d^2w/da^2$  
under the comparison with the numerical results and the observations.  The main essence of the allowed region for each potential is presented in Fig. 1.

 We have taken the exact approach \citep{11,12} to investigate five different exact potentials, not taken the approximate parametrization approach 
 for $w$ \citep{13,14}.  
We have also mainly concerned the terminal boundary condition at $z \simeq 0$ \citep {11,12,13,14,15,16,17,18,19,21,22,23} and not $ z \gg 1$ \citep{4,11,15,17}.  
Adding to the above-mentioned points, there are two more new aspects of this paper,

1) We have imposed a new constraint on $w(z)$, which was discovered in Ref. \citep{20} (Planck 2015).  
The main point of the new constraint is $w(z) \leq - 0.91 $
at $z = 0.65 \sim 2.0$, which has not been previously taken into consideration in the literature to our best knowledge.  
We consider that it is a handhold constraint to investigate the quintessence models.

2) We noticed that many investigated exact potentials, perhaps for simplicity, have two parameters and to determine these parameters from observations, we need $dw/da$ and $ d^2w/da^2$, adding $w_0$. 
By assuming two parameters $(dw/da,$ and $ d^2w/da^2)$, the potential parameters are fixed and we can calculate the evolution of scalar field $Q$ and variation of $w(z)$.
It becomes possible to compare the calculated variation of $w(z)$ with the observed constraint on $w(z)$.  
From this procedure, we can derive the allowed parameter space in $dw/da - d^2w/da^2$.  
The point is that the allowed region for each potential is different from each other and a coordinate point is a 
one-to-one correspondence to a set of potential parameters.  
In future, we hope that the more detailed direct observations of $dw/da-d^2w/da^2$ space become possible 
and they could distinguish the quintessence different potentials and the potential parameters.
So it must be the most fundamental step to understand the detailed dark energy in cosmology.

 In this paper, the allowed $dw/da-d^2w/da^2$ space was not obtained by direct observations 
of $dw/da$ nor $ d^2w/da^2$ but was obtained from the comparison of the observational constraint of $w(z \geq 0) $
 and the numerical simulations via scalar potentials, because the 
potential parameters were associated with $dw/da$ and $ d^2w/da^2$.

 In $\S 2$ the derivatives of $w$ with $a$ and the basic parameter $\Delta$ are presented.  The diagram and meaning of $dw/da-d^2w/da^2$ are introduced 
 and the numerical procedure with boundary conditions are explained.  The main results are presented in Fig. 1 and the followings are mainly the explanation for them. 
 In \S 3 the forbidden region for each potential is described and the allowed region is derived through the comparison with 
 the observations \citep{20}.  The detailed changes of $w(z)$ with the redshift $z$ for potentials and parameters in the $dw/da-d^2w/da^2$ space 
 are calculated.   In \S 4 the parameter $\Delta$ has been changed from 0.1 
 and the results of the allowed region for each potential are presented.  
 We investigate the effect of another observational constraint \citep{20} for the allowed regions for the potentials in $dw/da-d^2w/da^2$ space in \S 5.  
 In \S 6 the allowed regions for the investigated potentials are summarized and the possibility of the thawing and freezing type potentials are discussed. 
  Although it is calculated exactly in the text, it is approximated by Taylor expansion for the above potentials in Appendix A.

 \section{Derivatives of $w_Q$}
 
 \subsection{Scalar field}

For the dark energy, we consider the scalar field $Q$, where the action for this field in the gravitational field is described by  

 \begin{equation}
S=\int d^4 x \sqrt{-g} \left[ -\frac{1}{16 \pi G}R+ \frac{1}{2}g^{\mu \nu} \partial _ \mu Q \partial _ \nu Q -V( Q ) \right] +S_M ,  \label{act2}
\end{equation}
 where $S_M$ is the action of the matter field and $G$ is the gravitational constant, usually putting $G=1$ \citep{6}.
In the expanding universe, the equation for the scalar field $Q(t)$ becomes 
\begin{equation}
\begin{split}
\ddot{Q} + 3H \dot{Q}+V'=0  , \label{Qfield}
\end {split}
\end{equation}
where $H$ is the  Hubble parameter, overdot is the derivative with time, and prime of $V'$ is the derivative with $Q$.  
This is the equation to calculate the development of the field $Q$, if the potential is fixed with the estimated parameters.

The energy density and pressure for the scalar field are written by
\begin{align}
\rho _ Q =\frac{1}{2} \dot{Q}^2+V \label{density} ,   \end{align}
and
\begin{align}
p_ Q =\frac{1}{2} \dot{Q}^2-V  , \label{pressure}
\end{align}
respectively.  Then the parameter $w_Q$ for the equation of state is described by 
\begin{align}
w_ Q \equiv \frac{p_ Q}{\rho _ Q}=\frac{ \frac{1}{2} \dot{Q}^2-V}{ \frac{1}{2} \dot{Q}^2+V} =-1+\frac{\dot{Q}^2}{ \frac{1}{2} \dot{Q}^2+V} . \label{state}
\end{align}

It is assumed that the current value of $w_Q$ is slightly different from a negative unity by $\Delta ( > 0)$ 
 \begin{align}
w_Q=-1+\Delta     .\label{w}
\end{align}
By using Eq. (\ref{state}), $\dot{Q}^2$ is written as 
\begin{align}
\dot{Q}^2=\frac{\Delta V}{1-\frac{\Delta}{2}} ,  \label{(2)}
\end{align}
which becomes,
 \begin{align}
\dot{Q}^2=2(\rho_c\Omega_Q-V) , \label{(3)}
\end{align}
 where it is used the density parameter $\Omega_Q=\rho_Q/\rho_c$=0.68,  being $\rho_c=3H^2/(8\pi)$  the critical density of the universe.
Combining Eqs. (\ref{(2)}) and (\ref{(3)}), $V$ is given by
\begin{align}
V=\rho_c\Omega_Q\left(1-\frac{\Delta}{2}\right)=0.68\rho_c\left(1-\frac{\Delta}{2}\right) . \label{potential}
\end{align}
From Eqs. (\ref{(3)}) and (\ref{potential}), $\dot{Q}$ is expressed 
\begin{align}
\dot{Q}=\sqrt{\Delta(\rho_c\Omega_Q)}=0.825\sqrt{\rho_c\Delta}. \label{13}
\end{align}

Since $\rho_c$ is given by the observation through the Hubble parameter $H$, 
 $\dot{Q}$ is determined by  $\Omega_Q$ and $\Delta$, which also determine the value of $V$.  If we adopt the form and parameters of each potential, 
 the value of $V$ could be used to estimate the value of $Q$. 
 Actually, the evolution of $H$ in Eq. (\ref{Qfield}) depends on the background densities which include radiation density.  
 The effect of radiation density can be ignored in the near past ($z \leq 10^3$) and so is not considered in this work.  
 The numerical calculations must be done simultaneously with Eq. (2) and Friedmann Eq. for $\ddot{a}$ as
 \begin{align}
\frac{\ddot{a}}{a}=-\frac{4\pi}{3}\left( 3p_Q+ \rho _Q+ \rho_m \right) =-\frac{4\pi }{3}\rho_c((1+3w_Q)\Omega_Q+\Omega_m),  
\end{align}
 where $\rho_m$ is the matter density, including cold dark matter and baryonic matter.  
 We take $\rho_m(z)=\rho_{m0}(1+z)^3$, $\Omega_{m0}=\rho_{m0}/\rho_c=0.32$ and $\Omega_Q+\Omega_m=1$, taking $\rho_{m0}=\rho_m(z=0)$ and assuming flat universe.  
 It must be noted that if time $t$ is normalized by $t_n=(\sqrt{4\pi G \rho_c/3})^{-1}$ as $\tau=t/t_n$, it becomes $(da/d\tau)/a=\sqrt{2}$ at $a=1$ 
 for $H^2=((da/dt)/a)^2=8\pi G \rho_c/3$.

 \subsection{Diagram of $dw/da$ versus $d^2w/da^2$}
Figure 1 shows the summarized diagram of the derivatives of the equation of state $w(a)$ at the present time (scale factor $a$ = 1 and redshift $z$ = 0) 
for several scalar potentials :Inverse power law type (brown-curve like region), Exponential type (orange), Mixed type (yellow), 
cosine type (green), and Gaussian type (blue).  In Fig. 1, $d^2w/da^2$ are plotted against $dw/da$ for each scalar potential.  
A series of colored regions are the boundary and the allowed regions of $d^2w/da^2$ versus $dw/da$ which were obtained {\it a posteriori}, 
so as to make the numerical calculation $w(z)$ satisfy the observational constraint $w_{obs}(z) <$ $-$0.91 at $z$ = 0.65$-$2.0 \citep{20}, 
shown in Fig. 2 (b) by blue step lines.
  In the previous papers \citep{5, 6}, $dw/da$ and $d^2w/da^2$ were associated with two parameters of each scalar potential.  
When the fixed values of $w(z = 0)$, $\dot Q(z = 0)$, and $V(Q)(z = 0)$ are taken as the boundary conditions,
 each potential shape was found to be restricted in a some extent so as to reproduce the $w_{obs}(z)$ band data \citep{20}. 
 
  \subsection{Implications}
Figure 1 tells us that the allowed regions are almost mutually exclusive among the several potentials and that a coordinate point of ($dw/da$, $d^2w/da^2$) is nearly one-to-one correspondence to a set of the potential parameters. 
 According to the diagram of Fig. 1, the direct observation of ($dw/da$,  $d^2w/da^2$) enables us to identify not only the type of the scalar potential but also the potential parameters.   
The allowed regions in the $d^2w/da^2$ versus $dw/da$ diagram may be convenient and helpful to explore a likely scalar potential in quintessence models.

\begin{figure}[b]
\centering
\includegraphics[]{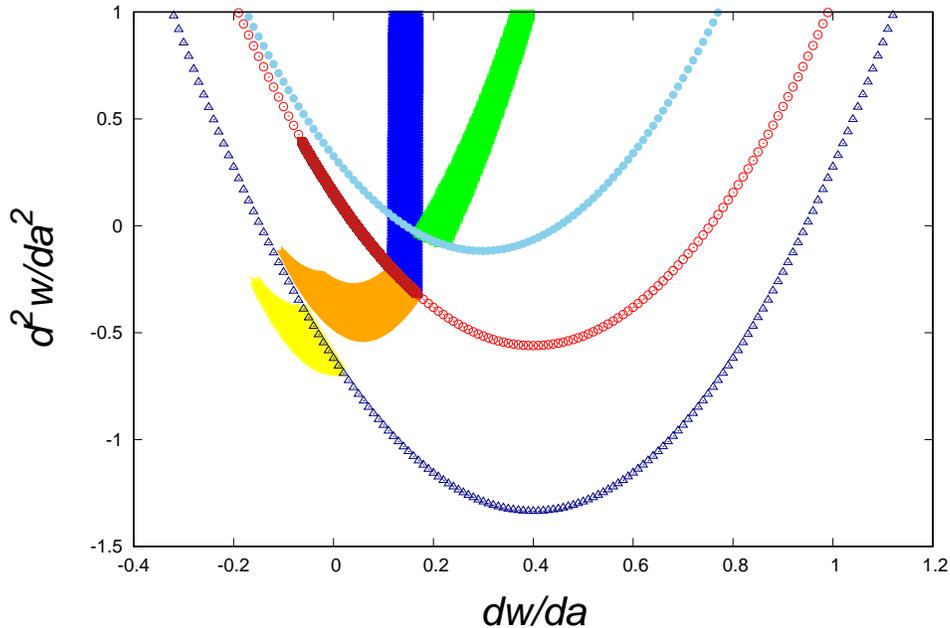}
\caption{$d^2w/da^2$ plotted against $dw/da$ illustrates the allowed regions for several scalar potentials : Inverse power law type (brown), 
Exponential type (orange), Mixed type (yellow), cosine type (green), and Gaussian type (blue).
A series of colored regions are the boundary and the allowed regions of $d^2w/da^2$ versus $dw/da$, 
so as to make the numerical calculation $w(z)$ with the unique boundary condition
 satisfy the observational constraint $w_{obs}(z) \leq $ $-$0.91 at $z$ = 0.65$-$2.0 \citep{20}.
The values of $dw/da$ and $d^2w/da^2$ were associated with two parameters of each scalar potential as implicit parameters.
One should note that the allowed regions are almost mutually exclusive and that a coordinate point is 
one-to-one correspondence to a set of potential parameters. }
\end{figure}
 
In the region $dw/da > 0$ of Fig. 1, $w$ is increasing at $z=0$ which shows a what is called "thawing feature". 
On the other hand, in $dw/da < 0$ , $w$ is decreasing at $z=0$ which shows a "freezing feature". 
In $d^2w/da^2 > 0$, $dw/da$ is increasing and $w(z)$ indicates a concave curve at $z=0$ which shows a "thawing feature" and a "freezing feature" \citep{18}. 
In $d^2w/da^2 < 0$, $dw/da$ is decreasing and $w(z)$ indicates a convex curve at $z=0$ which shows a "freezing feature". 

So "freezing feature" appears in the lower-left and upper-left regions on $dw/da - d^2w/da^2$ space and "thawing feature" appears in the upper-right region on $dw/da - d^2w/da^2$ space. 
In the other region, $w(z)$ shows a complex feature such as, for example in the right-lower region ($dw/da > 0$ and $d^2w/da^2 < 0$) where the initial conditions must be unnatural 
(to be discussed later).

  \subsection{Numerical procedure}
The procedure how to test the potential parameters and obtain the allowed regions in Fig. 1 is as follows. 
First, a set of values of ($dw/da$, $d^2w/da^2$) was chosen with a full plain scan, for example, of $-0.4 < dw/da < 1.2$ and  $-1.5 < d^2w/da^2 < 1.0$ by a step of 0.01. 
For each scalar potential, the two parameters were calculated as in the previous paper \citep{5} and touched in this paper. 
Second, a set of the terminal boundary conditions (given below) was imposed to solve the simultaneous differential equations of Eqs. (2) 
and (11) on the scalar field $Q$.
At the boundary condition, $Q(t)$ as a function of the time $t$ was numerically calculated. 
Then, the solutions of $\dot Q(t)$ and $V(Q)$ were obtained. 
These solutions were substituted into the equation of state $w(z)$ of Eq. (5).
Each $z$ dependence of $w(z)$ was numerically calculated by Eq. (5) according to the time development scale factor $a$, $\dot{Q}$ and $V(Q)$ for each potential.
Third, it was tested whether the numerical trajectory of $w(z)$ reproduces within the observational constraint of $w_{obs}(z)$ \citep{20}.
If yes, then the values of ($dw/da$, $d^2w/da^2$) are the allowed ones. Thus, the diagram was obtained.    

\subsection{Boundary conditions}
The first set of the terminal boundary conditions was imposed to solve the simultaneous differential equations of Eqs. (2) and (11) 
on the scalar field $Q$. 
The numerical solutions must satisfy the $z$ = 0 conditions of
\begin{align}
w(z = 0) = - 0.9, \nonumber   \\
\dot{Q(z = 0)}=\sqrt{\Delta(\rho_c\Omega_Q)}=\sqrt{(3\Delta \Omega_Q/(4\pi))}=0.127,  \nonumber  \\
V(Q)(z = 0)=\rho_c\Omega_Q\left(1-\frac{\Delta}{2}\right) =\frac{3}{4\pi}\Omega_Q\left(1-\frac{\Delta}{2}\right) = 0.154,  \nonumber
\end{align}
for the investigated potentials, where $\Delta=0.1$ and time normalization $t_n=(3/(4\pi  \rho_c))^{1/2}=1$ is taken.  
The initial value of $Q(z=0)$ is usually derived from $V(Q)(z=0)$, where the form parameters of each potential are already determined by 
adopted $dw/da$ and $d^2w/da^2$.  Under the boundary conditions, 
the potential parameters were restricted  to reproduce $w(z)$ within the observational constraint. 
A non-monotonic solution of $Q$ with respect to $z$ was found to be a key to cause the meandering behavior in $w_{obs}$. 
In $\S  4$, the effects of variation of $\Delta$ in $w$ = $-$1 + $\Delta$ were investigated.\\

\section{First and second derivatives of $w_Q$}
To fix the characteristic parameters of each potential, we use $dw_Q/da$ and $d^2w_Q/da^2$ \citep{2,5,6} as follows.  
The first derivative of $w_Q$ is given by
\begin{align}
\frac{dw_Q}{da} & = \frac{1}{\dot{a}}\frac{d}{dt}\left(\frac{p_Q}{\rho_Q}\right)=\frac{1}{\dot{a}}\frac{\dot{p}_Q\rho_Q-p_Q\dot{\rho}_Q}{\rho^2_{Q}} \nonumber \\
& = \frac{2V\dot{Q}}{aH\rho^2_Q}\left(-3H\dot{Q}-\frac{V'}{V}\rho_Q\right) , \nonumber
\end{align}
then $\frac{V'}{V}M_{\rm pl}$ is expressed by $dw_Q/da$ as 
\begin{align}
\frac{V'}{V}M_{\rm pl} = -\left(1-\frac{\Delta}{2}\right)^{-1}\sqrt{\frac{2\pi}{3\Delta\Omega_Q}}\left\{a\frac{dw_Q}{da}+6\Delta(1-\frac{\Delta}{2})\right\},
\label{X1}
\end{align}
where $M_{\rm pl}$ is the Planck mass.

From the form of $dw_Q/da$, the second derivative of $w_Q$  is given by
\begin{align}
\frac{d^2w_Q}{da^2} = \frac{1}{\dot{a}^3\rho_Q^4}[(\ddot{p}_Q\rho_Q-p_Q\ddot{\rho}_Q)\dot{a}\rho^2_Q-(\dot{p_Q}\rho_Q-p_Q\dot{\rho}_Q)(\ddot{a}\rho_Q^2+2\dot{a}\rho_Q\dot{\rho}_Q)].   
\label{d^2w/da^2}
\end{align}
After the calculation in the paper \citep{6}, $d^2w_Q/da^2 $ becomes 
\begin{align}
\frac{d^2w_Q}{da^2}=&  \frac{3}{4\pi}\frac{\Omega_Q}{a^2}\left(1-\frac{\Delta}{2}\right) \left[-\Delta M^2_{ \rm pl}\frac{V''}{V} 
+\sqrt{\frac{6\pi \Delta}{\Omega_Q}}\left((1-\Delta)(6+\Omega_Q)-\frac{1}{3} \Omega_Q \right)  \right. \nonumber  \\
& \left. \times M_{\rm pl}\left(\frac{V'}{V}\right) +\left(1-\frac{\Delta}{2} \right) M^2_{\rm pl} \left( \frac{V'}{V} \right)^2
+\frac{8\pi\Delta}{\Omega_Q}(7-6\Delta) \right] . 
\label{d^2wII}
\end{align}
From this equation, we estimate $d^2w_Q/da^2$ for each potential \citep{6}.  $(V''/V)M_{\rm pl}^2$ is expressed by $d^2 w_Q/da^2$ as
\begin{align}
\frac{V''}{V}M_{\rm pl}^2 = &-\frac{1}{\Delta} \left[\left(1-\frac{\Delta}{2}\right)^{-1}\frac{4\pi a^2}{3\Omega_Q}\frac{d^2 w_Q}{da^2}-
\sqrt{\frac{6\pi \Delta}{\Omega_Q}} \left \{ (1-\Delta)(6+\Omega_Q)-\frac{1}{3}\Omega_Q \right \} M_{\rm pl}\frac{V'}{V} \right.  \nonumber \\
& \left.-\left(1-\frac{\Delta}{2}\right)\left(M_{\rm pl}\frac{V'}{V}\right)^2 - \frac{8\pi}{\Omega_Q}\Delta(7-6\Delta) \right]. 
\label{Y2}
\end{align}
Thus $dw_Q/da$ and $d^2w_Q/da^2$ are associated with the parameters of potential form.  


Let us explain how to estimate the potential parameters and numerical simulations for three freezing type potentials of 
$ V=M^{4+\alpha}/Q^{\alpha}$ (inverse power law; $\alpha > 0$) \citep{8},
 $V=M^4\exp(\beta M/Q)$ (exponential; $\beta > 0$), and $V=M^{4+\gamma}/Q^{\gamma}\exp (4\pi Q^2)$ (mixed; $\gamma > 0$) \citep{9}. \\

\subsection{Inverse power law potential}


$\alpha$ is given by  
\begin{align}
\alpha=\left(\frac{V'}{V}\right)^2\left[\left(\frac{V''}{V}\right)-\left(\frac{V'}{V}\right)^2\right]^{-1}.
\label{ALPHA}
\end{align}

Using Eqs. (\ref{X1}), (\ref{Y2}), and (\ref{ALPHA}),
  $\alpha$ is described by $d^2w_Q/da^2$ and $dw_Q/da$ as
\begin{align}
\alpha =& -\left[2\Delta \left(1-\frac{\Delta}{2}\right)\right] ^{-1}\left(\frac{dw_Q}{da}
+6\frac{\Delta}{a}(1-\frac{\Delta}{2})\right)^{2}\left[\frac{d^2w_Q}{da^2} -  \hspace{3cm} \right. \nonumber \\
  &- \left. \frac{(1-3\Delta/2)}{2\Delta(1-\Delta/2)} \left(\frac{dw_Q}{da}- \frac{3\Delta(1-\Delta/2)}{a(1-3\Delta/2)}(1+\frac{(2-3\Delta)\Omega_Q}{6})\right)^2   
  \right. \nonumber \\
 &+ \left.  \frac{3\Delta(1-\Delta/2)}{8a^2(1-3\Delta/2)} \left(3((1-\Delta)(6+\Omega_Q)-\frac{\Omega_Q}{3})^2-16(7-6\Delta)(1-3\Delta/2) \right)\right]^{-1}.  
\label{alphaposi}
\end{align}
Then the constraint $\alpha > 0$ separates the region by the parabolic curve in Fig. 1 as 
\begin{align}
\frac{d^2w_Q}{da^2}&-\frac{(1-3\Delta/2)}{2\Delta(1-\Delta/2)} \left(\frac{dw_Q}{da}- 
\frac{3\Delta(1-\Delta/2)}{a(1-3\Delta/2)}(1+\frac{(2-3\Delta)\Omega_Q}{6})\right)^2  \nonumber \\
&+\frac{3\Delta(1-\Delta/2)}{8a^2(1-3\Delta/2)} \left(3((1-\Delta)(6+\Omega_Q)-\frac{\Omega_Q}{3})^2-16(7-6\Delta)(1-3\Delta/2) \right) <  0.
\end{align}

This parabolic criterion for $\Delta=0.1$ is shown by red curve in Fig. 1 and the downside region indicates $\alpha > 0$.

From the previous study \citep{5}
\begin{align}
\frac{V''}{V}=\frac{\alpha(\alpha+1)}{Q^2}, \ \ 
\frac{V'}{V}=-\frac{\alpha}{Q}, \ \ 
\left(\frac{V'}{V}\right)^2=\frac{\alpha^2}{Q^2}, 
\label{V2Valpha}
\end{align}

This potential has two parameters of $\alpha$ and $M$.  
$\alpha$ is estimated by $dw/da$ and $d^2w/da^2$ through Eq. (17).   $V'/V$ is estimated by $dw/da$ through Eq. (12).  
  Using these $\alpha$ and $V'/V$, $Q$ at present ($z$=0) is 
estimated through Eq. (19).  $M$ is given through 
$V=M^{4+\alpha}/Q^{\alpha}$ as $M=(Q^{\alpha}\rho_c\Omega_Q(1-\Delta/2))^{1/(4+\alpha)}$.

\begin{figure}[tb]
 \begin{minipage}[b]{.48\textwidth}
 \centering
\includegraphics[clip,width=7cm,height=5cm]{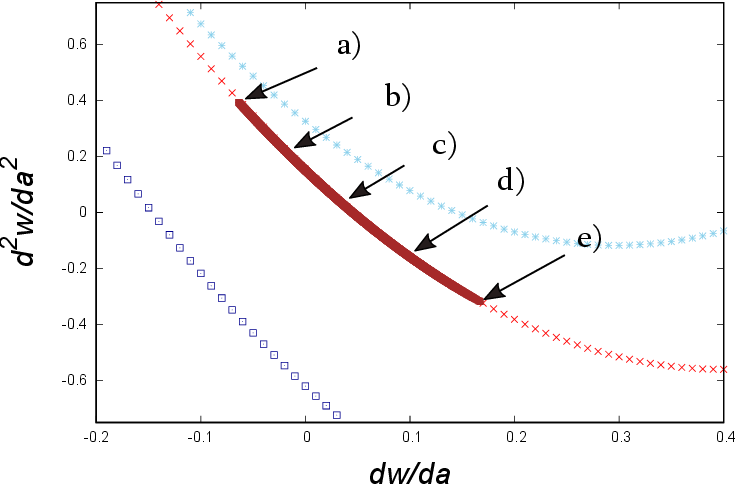}
 \subcaption{}
\label{fig:2a}
\end{minipage}
\hfill
\begin{minipage}[b]{.48\textwidth}
 \centering
 \includegraphics[clip,width=7cm,height=5cm]{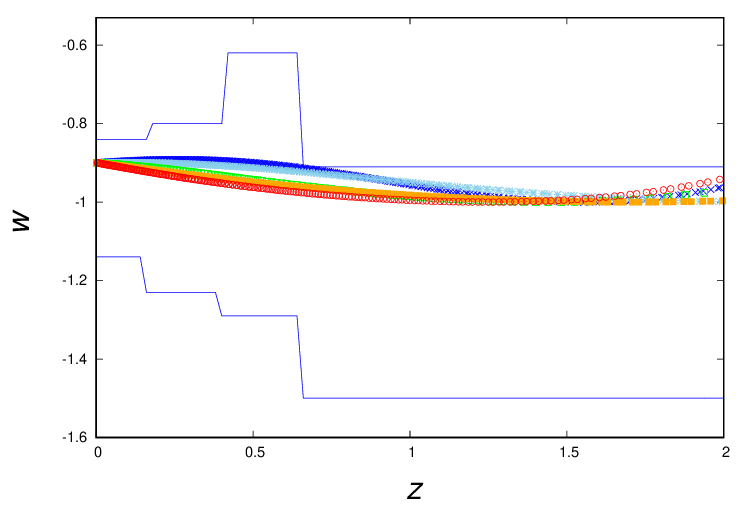}
 \subcaption{}
 \label{fig2b}
 \end{minipage}
\begin{minipage}[b]{.48\textwidth}
\centering
\includegraphics[clip,width=7cm,height=5cm]{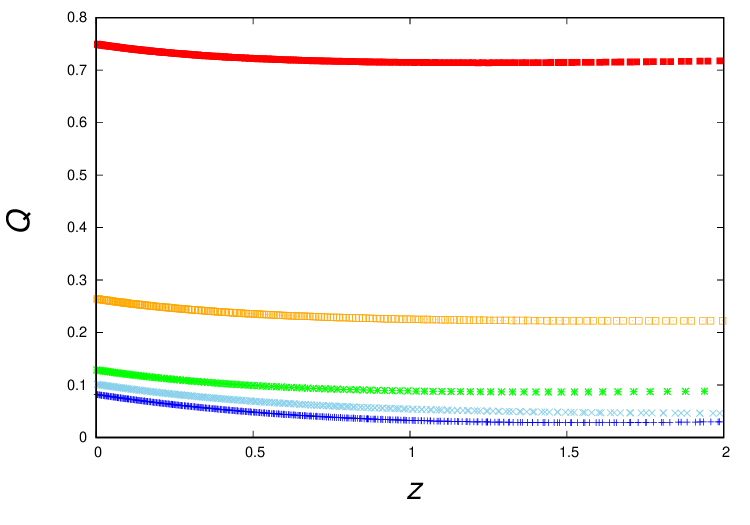}
\subcaption{}
\label{fig2c}
\end{minipage}
\hfill
 \begin{minipage}[b]{.48\textwidth}
\centering
\includegraphics[clip,width=7cm,height=5cm]{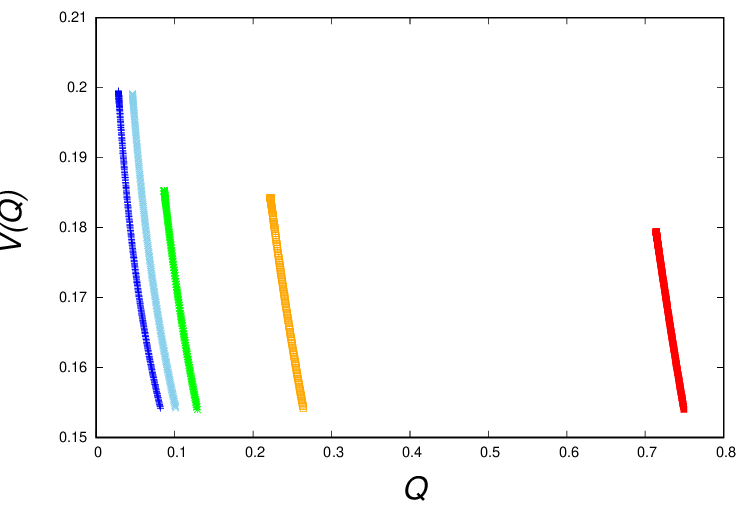} 
\subcaption{}
\label{fig2d}
\end{minipage}  
 \caption{Inverse power law potential.  (a) Brown streak-like region is the allowed region, enlarged Fig. 1.  
 The cases a) $\sim$ e) in Table I are indicated. 
  (b) Numerical calculations of $w$ as functions of $z$ with a) $\sim$ e).  
Blue step lines are the observational constraints for $w$.  $w$ must be in the narrow region ($-1 \le w \le 1$) for the evolution. 
 (c) Numerical calculations of $Q$ as functions of $z$ from the past ($z=2$) to the present ($z=0$) for the cases a) $\sim$ e).   
 (d) Falling down of scalar field $Q$ from the past ($z=2$) to the present ($z=0$).}
\end{figure}

By taking $\dot{Q}$ estimated by Eq. (10), the initial conditions for $Q$ are derived.    
By calculating Friedmann Eq. (11) for $a$ simultaneously, it is possible to estimate 
the evolution of $Q$ through Eq. (2) under the potential $V$.  Through the development of $Q$, 
it becomes possible to estimate the evolution of $w$.

In Fig. 2 (a), the allowed region in $dw/da-d^2w/da^2$ is presented, where the allowed typical cases a) $\sim$ e) are indicated.
The cases a) $\sim$ e) with values of  $\alpha$ and $Q_0$ are presented in Table I. 
 The observational constraints by blue step lines and numerical calculations of $w$ as a functionof $z$ for the cases a) $\sim$ e) 
 are shown in Fig. 2 (b).  $Q$ as a functionof $z$ is presented in Fig. 2 (c).  In Fig. 2 (d), 
 $V$ as a functionof $Q$ is shown where $z$ is an implicit parameter.

In Fig. 3, the evolutions for typical cases of a) $\sim$ e) are presented for $V(Q)$ versus $z$ in (a), $dQ/dt$ versus $z$ in (b), 
$V$ versus $dQ/dt$, and $V$ versus $d^2Q/dt^2$, respectively.
The evolutions for typical cases of a) $\sim$ e) are presented in Fig. 4 for $w$ versus $z$ in (a), $d^2a/dt^2$ versus $z$ in (c), 
and $da/dt$ versus $z$ in (d), respectively.  
It is shown in Fig. 4 (b) by red curves the $w$ evolution of almost 100 cases evenly adopted from the brown region in Fig. 2 (a).

\begin{table}[htb]
\centering
  \begin{tabular}{|l|r|r|r|r|r|} \hline
       & $dw/da$      &      $d^2w/da^2$ &      $\alpha$ & $Q_0$  & color \\ \hline \hline
    a) & $-6.37\times10^{-2}$ &    0.394 & 0.248& $8.38\times10^-2$           & blue   \\ \hline
    b) &   $-2.00\times10^{-2}$ &    0.224 & 0.340& 0.101           & sky-blue \\ \hline
    c) & $ 5.01\times10^{-2}$ &$-1.55\times10^{-2}$ & 0.466& 0.129 & green \\ \hline
    d) &     0.10     & -0.159           & 1.033     &0.263      & orange \\ \hline
    e) &     0.15     & -0.281           & 3.15       & 0.749     & red \\ \hline 
  \end{tabular}
  \caption{The allowed values of $dw/da$ and $d^2w/da^2$ for typical cases of the inverse power law potential, 
  and the estimated $\alpha$ and $Q$ at present ($Q_0$).}
\end{table}


\begin{figure}[H]
\begin{minipage}[b]{.48\textwidth}
\centering
\includegraphics[clip,width=7cm,height=5cm]{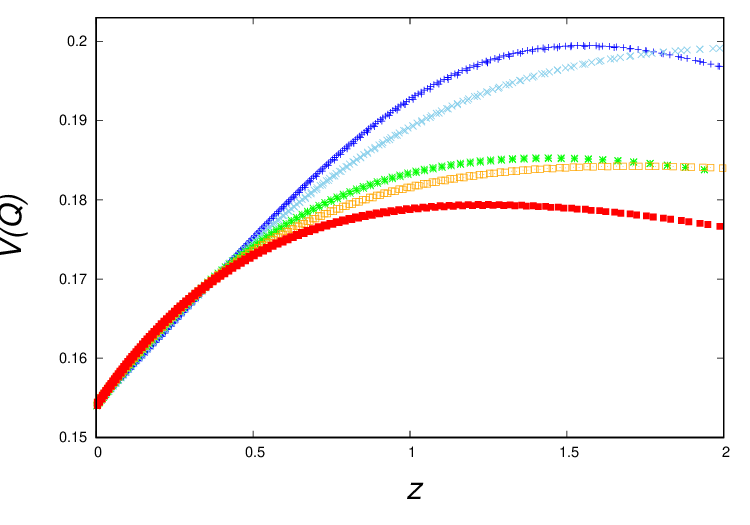}
\subcaption{}
\label{fig3c}
\end{minipage}
\hfill
 \begin{minipage}[b]{.48\textwidth}
\centering
\includegraphics[clip,width=7cm,height=5cm]{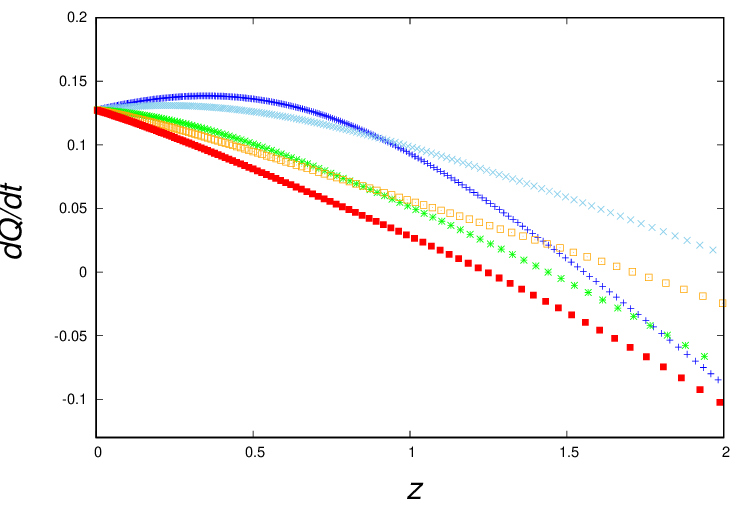}
\subcaption{}
\label{fig3d}
\end{minipage} 
\begin{minipage}[b]{.48\textwidth}
\centering
\includegraphics[clip,width=7cm,height=5cm]{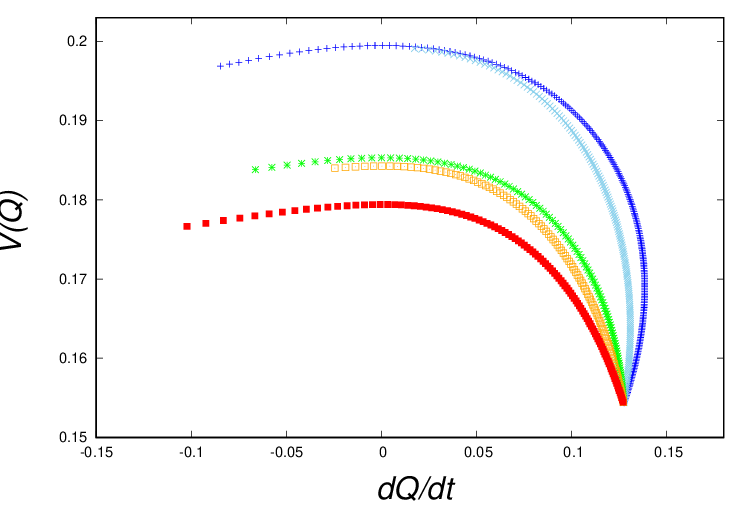}
\subcaption{}
\label{fig3e}
\end{minipage}
\hfill
 \begin{minipage}[b]{.48\textwidth}
\centering
\includegraphics[clip,width=7cm,height=5cm]{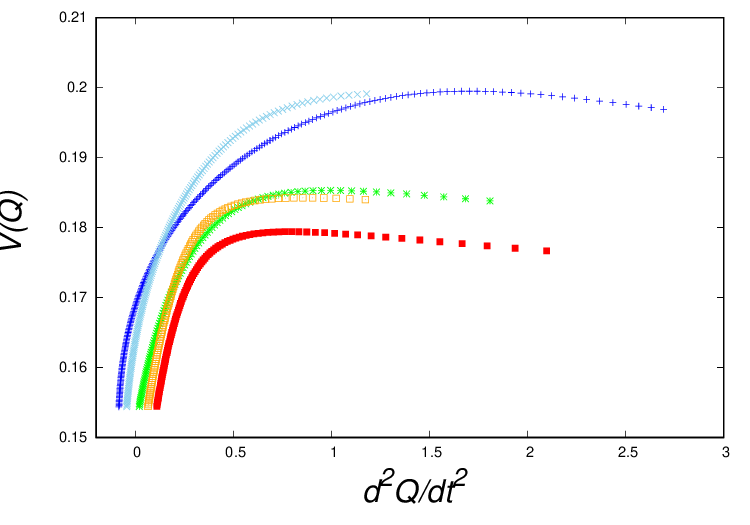}
\subcaption{}
\label{fig3f}
\end{minipage} 
 \caption{Numerical calculations of time evolution of $V(Q)$ (a) and $dQ/dt$ (b) as functions of $z$.  
 $V(Q)$ is plotted against $dQ/dt$ (c) and $d^2Q/dt^2$ (d) with $z$ as an implicit parameter.  
 Confront color sorting with Table I but not Fig. 1.}
 \end{figure}
 


The observational constraint in Fig. 2 (b) is taken from Fig. 7 in the reference (Ade. et al. \citep{20}), 
which includes the Planck 2015 results and 
BSH (Baryonic Oscillation Effect, Supernova Ia and Hubble parameter).  
By adopting the values of $dw/da$ and $d^2w/da^2$ at $z=0$, 
the characteristic parameters  such as $\alpha$ and $M$ are fixed and then the time development of $Q$ can be calculated.  
If the time variation of $w(z)$  does not conflict with the observational constraint, the adopted values of $dw/da$ and $d^2w/da^2$ are allowed.  
Namely, the observational constraint determines the allowed region in $dw/da-d^2w/da^2$ space.

The allowed region for the potential is in the brown curve-like in Figs. 1 and 2 (a).  The typical evolutional cases in Table I are 
presented in Figs. 2, 3, and 4.  The curves of $w$ evolution are reversible.  
The most severe constraint for the potential 
parameters ($dw/da$ and $d^2w/da^2$) comes from the observational upper limit line of $z \geq 0.65$, where $w \leq -0.91$, 
which could be understood in Figs. 4 (a) and (b).

The cases (marked by a) and b)) in the left-hand part of the brown region ($dw/da < 0  $) in Fig. 2 (a) 
are represented in Figs. 2, 3 and 4 by blue and sky-blue curves of which slopes are positive at $z=0$ ($dw/dz=-(1+z)^{-2}dw/da > 0$ ).  
The second derivative signature of $w$ with $z$ is not necessary the same with the second derivative signature of $w$ with $a$ as $
\frac{d^2w}{dz^2}=-\frac {d}{dz} \left( \frac {1}{(1+z)^2} \right ) \frac{dw}{da} -\frac{1}{(1+z)^2}\frac{da}{dz}\frac{d^2w}{da^2}
 =\frac{2}{(1+z)^3}\frac{dw}{da}+\frac{1}{(1+z)^4}\frac{d^2w}{da^2}.$  In this left-hand part of the brown region, 
 the signature of $d^2w/dz^2$ is negative and they are the curves of convex upward which are represented by blue, and sky-blue color curves.
   Especially the case a) represented by blue curve shows that $w$ has decreased from upper value to $-0.9$ at present ($z=0$) which means $\dot{Q}^2$ has decreased in the potential $V$.  
The freezing potential has the feature that $w$ will decrease and approach to $-1$.

The cases  (marked by c), d), and e))  in the right-hand part of the brown region ($dw/da > 0$) in Figs. 1 and 2 (a) are represented 
in Figs. 2, 3 and 4 by green, orange, and red curves. 

\begin{figure}[H]
 \begin{minipage}[b]{.48\textwidth}
\centering
   \includegraphics[clip,width=7cm]{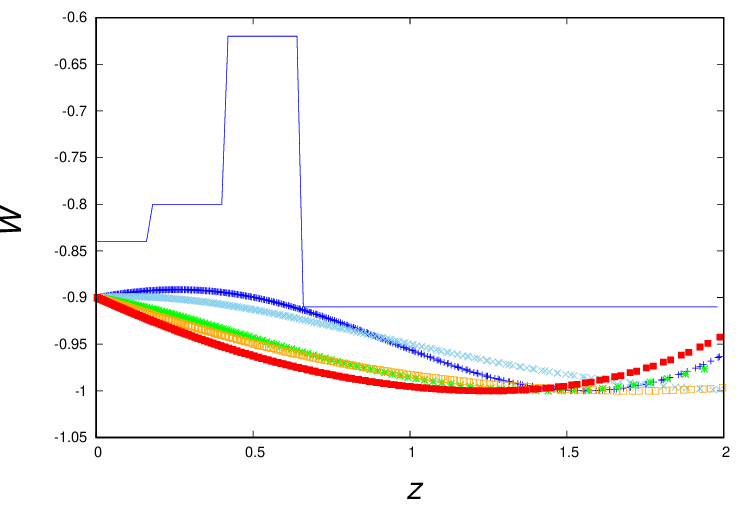}
  \subcaption{}
  \label{fig:4a}
 \end{minipage}   
\hfill
 \begin{minipage}[b]{.48\textwidth}
\centering
   \includegraphics[clip,width=7cm]{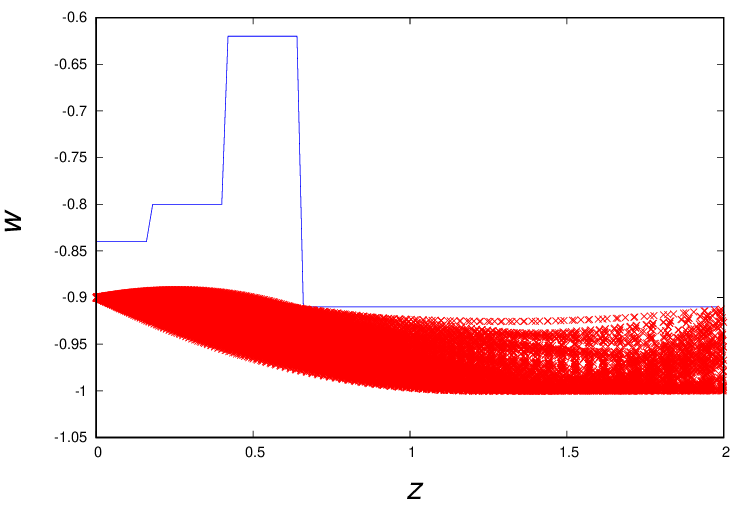}
  \subcaption{}
  \label{fig:4b}
 \end{minipage} 
 \begin{minipage}[b]{.48\textwidth}
\centering
   \includegraphics[clip,width=7cm]{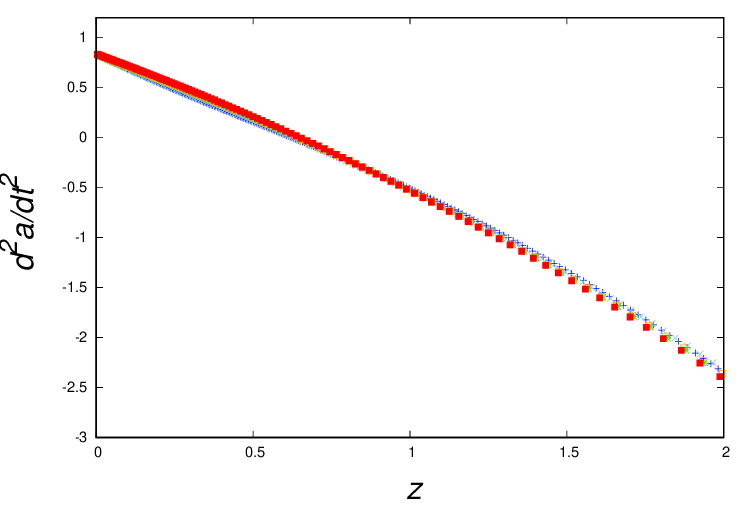}
  \subcaption{}
  \label{fig:4c}
 \end{minipage}   
\hfill
 \begin{minipage}[b]{.48\textwidth}
\centering
   \includegraphics[clip,width=7cm]{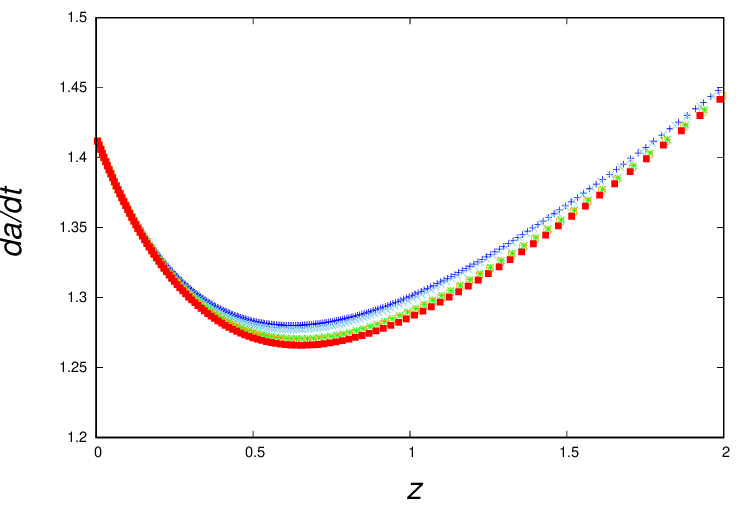}
  \subcaption{}
  \label{fig:five}
 \end{minipage} 
 \caption{Numerical calculations of time evolution of $w$ (a), $d^2a/dt^2$ (c), and $da/dt$ (d) for cases a) $\sim$ e) are presented.
About 100 cases are shown which are adopted from the brown region in Fig. 2 (a).}
\end{figure}
\hspace{-0.5cm}of which slopes are negative at $z=0$ ($dw/dz=-(1+z)^{-2}dw/da < 0$ ).  In these cases $w$ has increased from $-1$ to $-0.9$ 
which means that $\dot{Q}^2$ has increased in the potential.  
The thawing type potential has this feature.  The second derivative signature of $w$ with $z$ is positive and they are the curves of downward convex (concave).

 In Fig. 3 (a), it can be seen that all cases except b) in Table I have climbed up the potential at first and then passed down.
 All cases except b) have begun from a negative value of $dQ/dt$ and then increased in Fig. 3 (b). 
 When $dQ/dt$ passes through $0$, $w$ takes the value of $-1$  which is noticed in Fig. 4 (a).     
 The blue a) and sky-blue b) cases have decreased the value of $dQ/dt$ after $z \sim 0.5$ which can be seen in Fig. 3 (b). 
  It may be due to the expansion drag effect of the second term in Eq. (2).
  (c) Relation between $dQ/dt$ and $V(Q)$ for each case. 
  In Fig. 3 (c), it can be seen that $dQ/dt$ has increased for almost all cases,  however it has decreased (after $z \sim 0.5$) 
  for the blue a) and sky-blue b) cases.  (d) Relation between $d^2Q/dt^2$ and $V(Q)$ for each case.  
  It can be noticed in Fig. 3 (d) that $d^2Q/dt^2$ is positive in almost all situation, 
  however it becomes negative (after $z \sim 0.5$) for the blue a) and sky-blue b) cases.

In Fig. 4 (a), under the observational constraint, the $w$ evolution of cases a) $\sim$ e) are shown, which are the same as Fig. 2 (b), 
except the part of $-1.05 \le w \le -0.5$ in the vertical axis is enlarged.  
The a) and b) cases in the left-hand part of the brown region ($dw/da < 0$) in Fig. 2 (a) are represented in Fig. 4 a) by blue 
and sky-blue curves of which slopes are positive at z=0 \ ($dw/dz=-(1+z)^{-2}dw/da > 0$). 
In Fig. 4 (b), almost 100 cases are presented by red curves which are evenly adopted from the brown region in Fig. 2 (a). 
Evolutions of $d^2a/dt^2$ for cases a) $\sim$ e) are presented in Fig. 4 (c) where the signature has changed to positive around $z=z_a \sim 0.6$, 
meaning to become accelerating universe.  The fact that $z_a$ is so close to zero is the coincidence problem.  
In Fig. 4 (d), the evolution of $da/dt$ are shown.  Expansion velocity was decreasing at first and then changed to increasing at around 
  $z \sim 0.6$.  These changes of Figs. 4 (c) and (d) are almost the same for the following cases in Tables II $\sim$ V, 
  so they are not shown anymore.


Even for the freezing type potential, there are some cases which have the thawing features that $w$ has increased from lower value $(\simeq -1)$ to $-0.9$.    
 These characteristics are common for the following exponential and mixed type potentials.  

 \subsection{exponential potential}
 The constraint $\beta > 0$ is given from the relation

 \begin{align}
 \beta=\frac{2Q}{M}\left(\frac{V'}{V}\right)^2\left[\left(\frac{V''}{V}\right)-\left(\frac{V'}{V}\right)^2\right]^{-1}.
 \label{BETA}
 \end{align}
Taking $ Q > 0$, the criterion of $\beta > 0$ and $\alpha > 0$ \  is the same as the positive signature of 
the square brackets [\ \ \ \ ] and 
the same region in 
$dw_Q/da-d^2w_Q/da^2$ plane of Fig. 1.

The cases a) $\sim$ e) with values of $dw/da, d^2w/da^2, \beta M$ and $Q_0$ are presented in Table II.  
In Fig.5 (a), it is the same as Fig. 1 for the allowed orange colored region for the exponential potential which is enlarged.
 The cases a) $\sim$ e) in Table II are indicated there.    
 In Fig. 5 (b), downward movement of the  field $Q$ for each case a) $\sim$ e)  are shown from $z=2$ to $z=0$.  
 For each case, the parameter $\beta$ is not the same, so the form of the potential
  $V=M^4\exp(\beta M/Q)$ is mutually different. 
 However the value of potential $V$ at bottom is the same at present for every case as 
 $V_0=\rho_c\Omega_Q(1-\Delta/2) \simeq 0.154$ (Eq. (9)).  
   Evolution of the potential height $V(Q)$ for each case is shown in Fig. 5 (c)
 Three cases ( a), c) and e)) have climbed the potential at first ($z \sim 2$) and then gone down.  
 When $dQ/dt$ passes through $0$, $w$ takes the value of $-1$  which is noticed in Fig. 6 (a). 
 In Fig. (d), the evolution of the velocity $dQ/dt$ for each case is presented. 
 Three cases ( a), c) and e)) have started from $dQ/dt < 0$ at first (climbed the potential) 
 and then increased to become positive.

The observational constraints by blue step lines and numerical calculations of $w$ as a function of $z$ for the cases a) $\sim$ e 
 are shown in Fig. 6 (a).  Many cases in the orange region in Fig. 5 (a) are shown by red curves in Fig. 6 (b)

For this potential, the allowed parameter region in $dw_{Q}/da-d^2w_{Q}/da^2$ plane is presented by orange color in Figs. 1 and 5 (a).  
The example curves in Table II are shown in Figs. 5 and  6 (a).
The procedure of calculation is almost the same with the inverse power law potential.  
Taking $dw/da$ and $d^2w/da^2$, $Q$ is derived by Eq. (42) in the reference (\citep{5}) and so on (see Appendix B).
For the following potentials, the detailed calculations are explained in the Appendix B and the reference (\citep{5}).

\begin{table}[htb]
\centering
  \begin{tabular}{|l|r|r|r|r|r|} \hline
       & $dw/da$      &      $d^2w/da^2$ &      $\beta M$ & $Q_0$   & color \\ \hline \hline
    a) & -0.10441 &   -0.11599 & $3.98\times10^{-2}$  & 0.121   & blue \\ \hline
    b) &   $-1.83\times10^{-2}$ &  -0.232 & 0.154 & 0.219           & sky-blue \\ \hline
    c) & $ 5.06\times10^{-2}$ & -0.522 & 0.177 & 0.220        & green \\ \hline
    d) &     0.114     & -0.204           & 552   & 11.8      & orange \\ \hline
    e) &     0.168     & -0.332           & 546   & 11.3       & red \\ \hline 
  \end{tabular}
  \caption{The values of $dw/da$ and $d^2w/da^2$ for typical cases of the exponential potential in Figs. 5 and 6 (a).  
  The cases of a) to e)  are adopted from the boundary of the orange region in Figs. 1 and 5 (a).  
  It should be noted that the parameter $\beta M$ and the value of $Q_0$ at present are different for each case.}  
\end{table}

    
\begin{figure}[H]
\begin{minipage}{.48\textwidth}
\centering
   \includegraphics[clip,width=70mm,height=5cm]{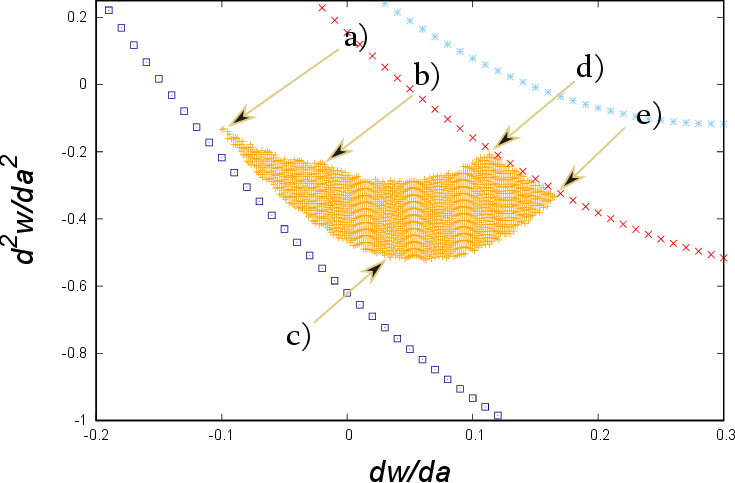}
  \subcaption{}
  \label{fig5a}
 \end{minipage}
 \hfill
 \begin{minipage}{.48\textwidth}
\centering
   \includegraphics[clip,width=70mm,height=5cm]{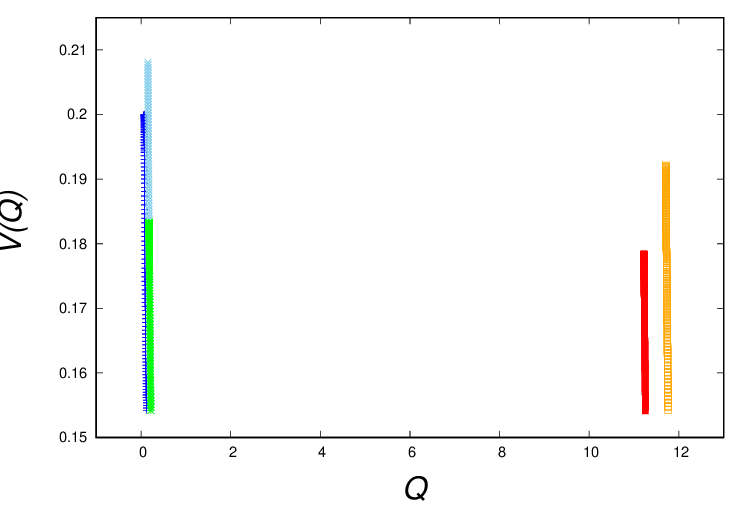}
  \subcaption{}
  \label{fig5b}
 \end{minipage}
 \begin{minipage}{.48\textwidth}
\centering
   \includegraphics[clip,width=70mm,height=5cm]{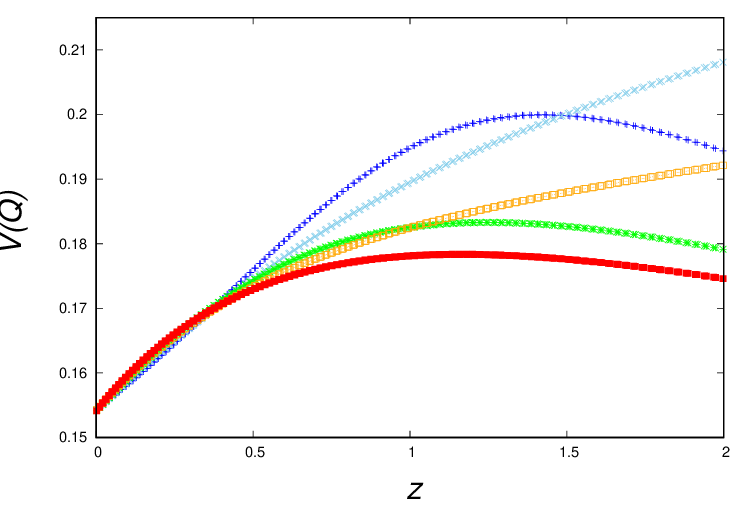}
  \subcaption{}
  \label{fig5c}
 \end{minipage}
 \hfill
 \begin{minipage}{.48\textwidth}
\centering
   \includegraphics[clip,width=70mm,height=5cm]{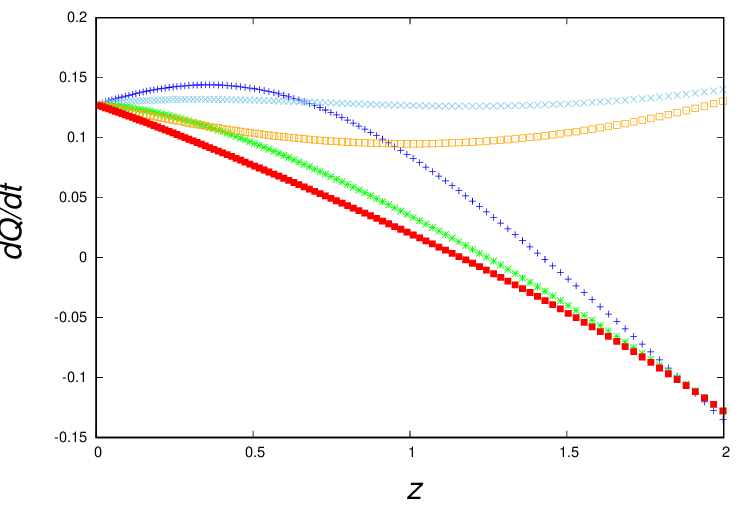}
  \subcaption{}
  \label{fig5d}
 \end{minipage}
 \caption{Exponential potential.  The allowed region is presented (a), where the typical cases a) $\sim$ e) are indicated.
$V$ versus $Q$ is presented in Fig. 5 (b), where $z$ is the implicit parameter.   
$V$ and $dQ/dt$ versus $z$ are presented in Fig. 5 (c) and (d). }   
\end{figure}

\begin{figure}[b]
 \begin{minipage}{.48\textwidth}
\centering
   \includegraphics[clip,width=70mm]{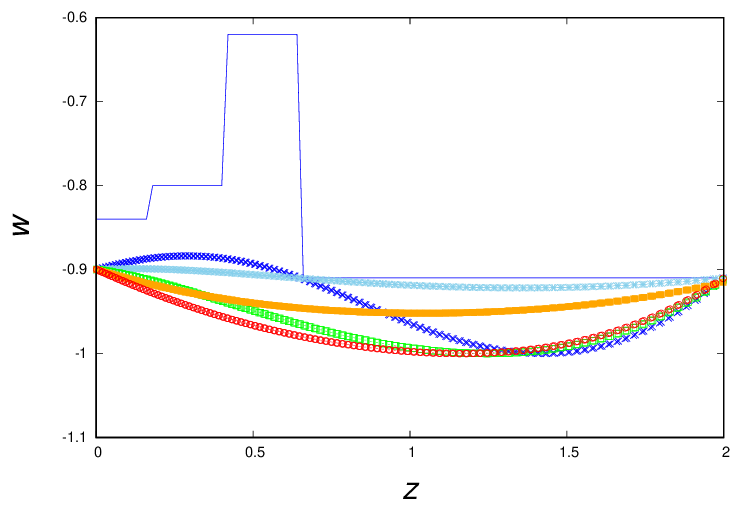}
  \subcaption{}
  \label{fig:6a}
 \end{minipage}
 \hfill
 \begin{minipage}{.48\textwidth}
\centering
   \includegraphics[clip,width=70mm]{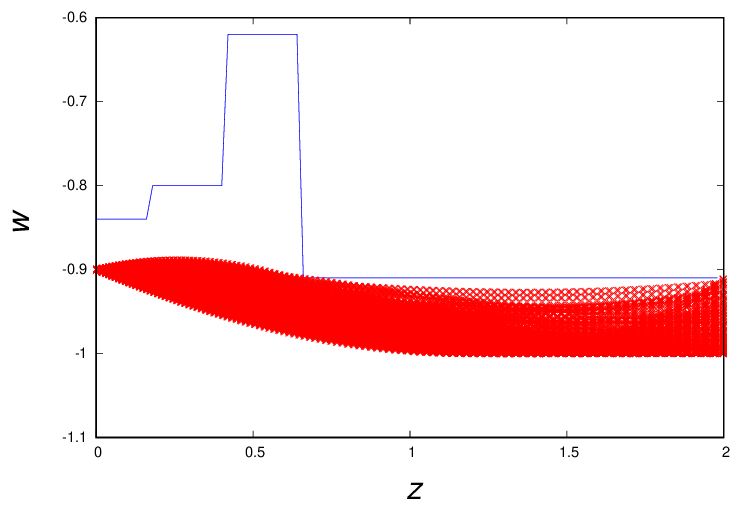}
  \subcaption{}
  \label{fig:6b}
 \end{minipage}
 \caption{(a) Under the observational constraint of blue step lines, the allowed typical cases are displayed.  
 (b) Many cases in the orange region in Fig. 5 (a) are shown by red curves.}
\end{figure}

\subsection{Mixed type potential}

The constraint $\gamma > 0$ is given by
   \begin{align}
  \gamma=\left(\frac{V'}{V}\right)^2\left[\left(\frac{V''}{V}\right)-\left(\frac{V'}{V}\right)^2-8\pi \right]^{-1}.
   \end{align}
   This relation is reduced to
  
  \begin{align}
\frac{d^2w_Q}{da^2}&-\frac{(1-3\Delta/2)}{2\Delta(1-\Delta/2)} \left(\frac{dw_Q}{da}- \frac{3\Delta(1-\Delta/2)}{a(1-3\Delta/2)}(1+\frac{(2-3\Delta)\Omega_Q}{6})\right)^2  \nonumber \\
&+  \frac{3\Delta(1-\Delta/2)}{8a^2(1-3\Delta/2)} \left(3((1-\Delta)(6+\Omega_Q)-\frac{\Omega_Q}{3})^2-16(7-6\Delta-\Omega_Q)(1-3\Delta/2) \right) < 0.
\end{align}
 which is almost to the same of the $ \alpha  \ > 0 $  constraint (Eq. (19)), however the last constant term is different from $16(7-6\Delta)(1-3\Delta/2) $ to  
 $16(7-6\Delta-\Omega_Q)(1-3\Delta/2)$.  This criterion is shown in Fig. 1 by dark-blue parabolic curve.

For this potential, the allowed parameter region in $dw_{Q}/da-d^2w_{Q}/da^2$ plane is shown in yellow in Figs. 1 and 7 (a).  
The typical cases a) $\sim$ e) with values of $dw/da, d^2w/da^2, \gamma,$ and $Q_0$ are presented in Table III and their calculated curves are shown in Figs. 7 and 8.  
The slopes of almost all lines are negative ($dw/da \leq 0$) which is understandable for the positive slope in $dw/dz \ (=-(1+z)^{-2} dw/da \geq 0)$ in Fig. 8. 
The allowed region is negative on the vertical axis ($d^2w/da^2 < 0$) which are the curves of convex upward shape around $z=0$.

 
 
 \begin{table}[htb]
\centering
  \begin{tabular}{|l|r|r|r|r|r|} \hline
       & $dw/da$      &      $d^2w/da^2$ &      $\gamma$ & $Q_0$   & color \\ \hline \hline
    a) & -0.163 &   -0.238 & 0.603 & 0.115   & blue \\ \hline
    b) & -0.111 &  -0.458 & 0.893 & 0.143           & sky-blue \\ \hline
    c) & $ -6.04\times10^{-2}$ & -0.395 & 12.7 & 0.653        & green \\ \hline
    d) &  $-2.96\times10^{-2}$  & -0.654           & 3.96   & 0.339     & orange \\ \hline
    e) &  $1.66\times10^{-2} $   & -0.682           & 5314   & 14.5      & red \\ \hline 
  \end{tabular}
  \caption{The values of $dw/da$, $d^2w/da^2$, $\gamma$, and $Q_0$ for typical cases of the mixed type potential in Figs. 7 and 8 (a).  
  The cases of a) to e)  are adopted from the boundary of the yellow region in Figs. 1 and 7 (a). }  
\end{table}


\begin{figure}[H]
 \begin{minipage}{0.48\textwidth}
  \centering
   \includegraphics[clip,width=70mm]{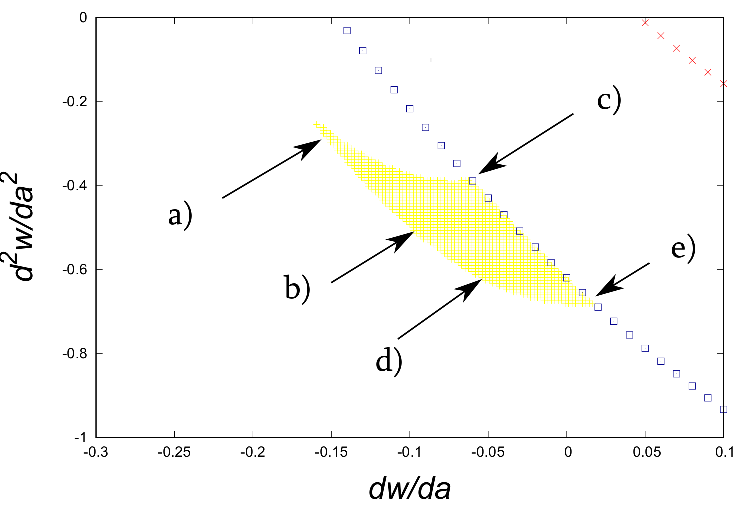}
  \subcaption{}
    \label{fig:7a}  
 \end{minipage}
 \hfill
 \begin{minipage}{0.48\textwidth}
  \centering
   \includegraphics[clip,width=70mm]{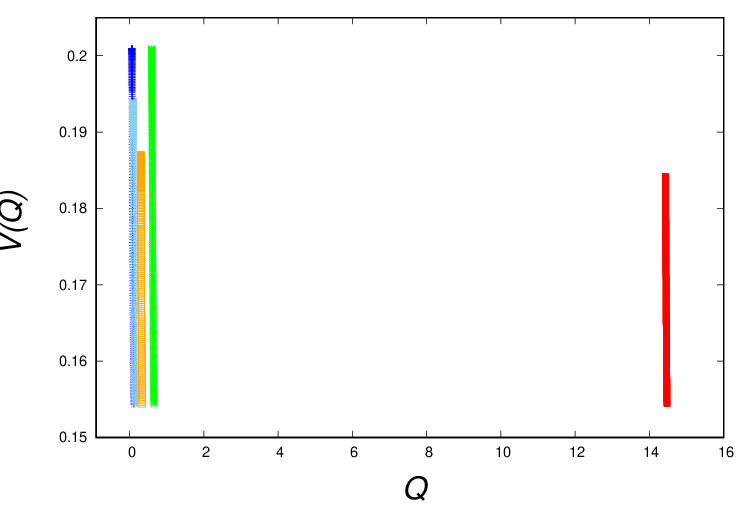}
  \subcaption{}
  \label{fig:7b}
 \end{minipage}
  \begin{minipage}{0.48\textwidth}
  \centering
   \includegraphics[clip,width=70mm]{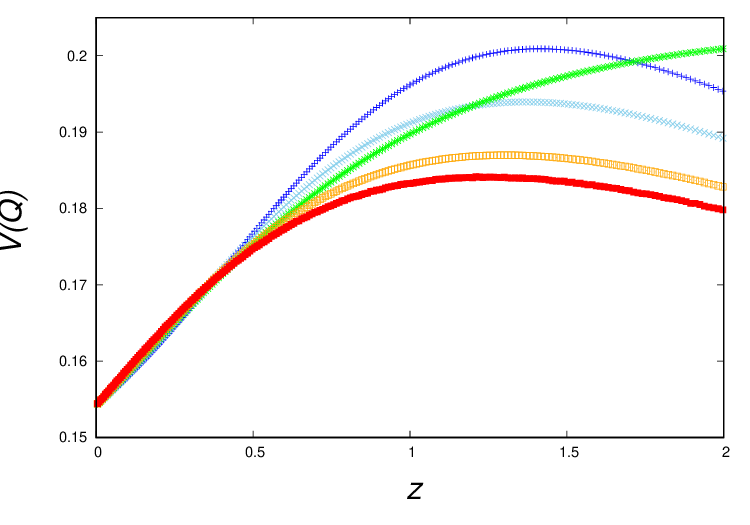}
  \subcaption{}
    \label{fig:7c}  
 \end{minipage}
 \hfill
 \begin{minipage}{0.48\textwidth}
  \centering
   \includegraphics[clip,width=70mm]{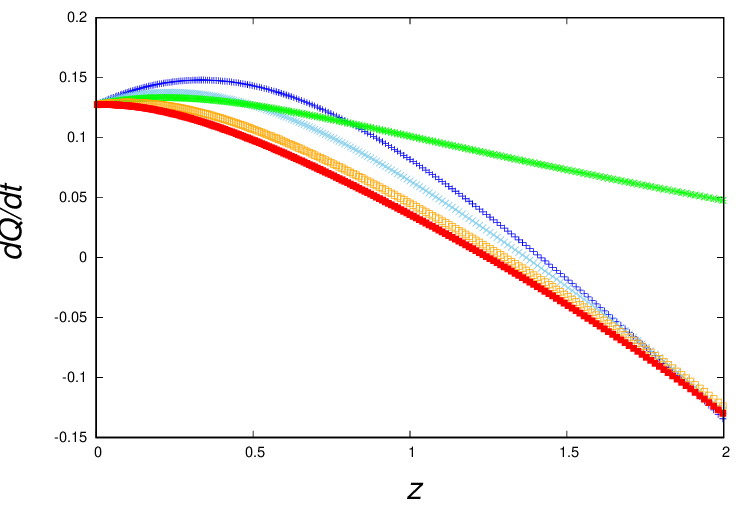}
  \subcaption{}
  \label{fig:7d}
 \end{minipage}
 \caption{Mixed type potential.  (a) The same with Fig. 1 for the allowed yellow  colored region for the mixed type potential which is enlarged.  
 The cases a) $\sim$ e) in Table III are indicated. (b) Downward movement of the field $Q$ from $z=2$ to $z=0$ in the mixed type potential.  
 For each case, the parameter $\gamma$ is not the same, so the forms of the potential
  $V=M^{4+\gamma}\exp(4\pi Q^2)/Q^{\gamma}$ are different. 
 However the value of potential $V$ at bottom is the same at present for every case as 
 $V_0=\rho_c\Omega_Q(1-\Delta/2) \simeq 0.154$ (Eq. (9)).  
 (c) Evolution of the potential height $V(Q)$ for each case is shown.  
 All cases except c) have climbed at first ($z \sim 2$) and then rolled down.
 (d) Evolution of the velocity $dQ/dt$ for each case is presented.  All cases except c) have started from $dQ/dt < 0$ at $z=2$ and then become positive. 
  When $dQ/dt$ passes through $0$, $w$ takes the value of $-1$  which is noticed in Fig. 8 (a). }   
\end{figure}

\begin{figure}[htbp]
 \begin{minipage}{0.48\textwidth}
  \centering
   \includegraphics[clip,width=70mm]{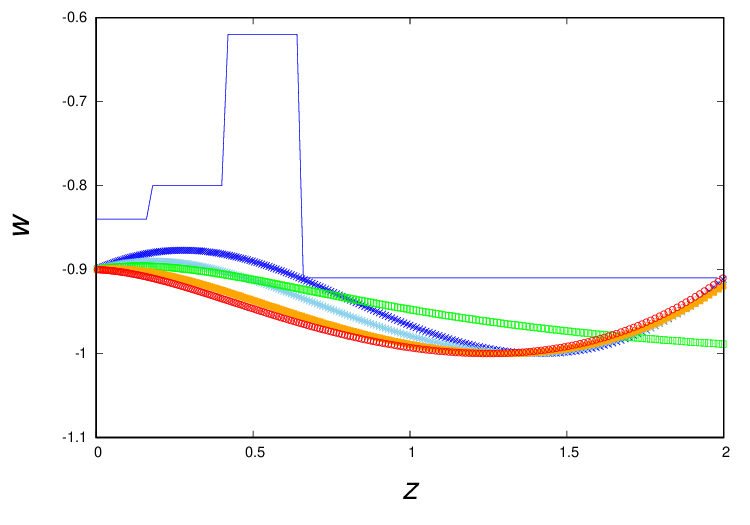}
  \subcaption{}
  \label{fig:8a}  
 \end{minipage}
 \hfill
 \begin{minipage}{0.48\textwidth}
  \centering
   \includegraphics[clip,width=70mm]{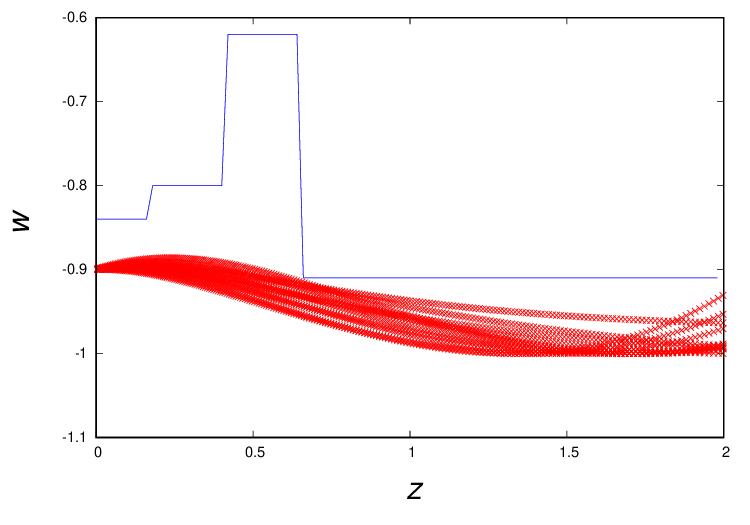}
  \subcaption{}
  \label{fig:8b}
 \end{minipage}
 \caption{(a) The constraint from the observation is shown by blue lines.  Under these constraint, 
the allowed typical curves for the mixed type potential are displayed. (b) Under the observational constraint, 
the allowed almost all cases for the mixed type potential in the yellow region in Figs 1 and 7 (a) are displayed by red curves. }
\end{figure}


The minimum point of this potential is realized at $Q_{min}=(\gamma/(8\pi))^{1/2}$. 
The values of $Q_0/Q_{min}$ for $a) \sim e)$ are  0.742, 0.761, 0.918, 0.854, and $\sim 1$, respectively.  
The field $Q$ has almost fallen to the minimum of the potential fot the case e).  As $\Delta=\dot{Q}^2/(\dot{Q}^2/2+V)= 0.1$, the field $Q$ has not yet stopped.  
The other cases are still rolling down the potential at present.

  \subsection{Cosine type potential}

For thawing potential, we investigate two type potentials such as  $V=M^4(\cos (Q/f)+1))$ (cosine)  and $V=M^4\exp(-Q^2/\sigma^2)$ (Gaussian). 
   In this thawing model, the field is assumed to be nearly constant at first and then starts to evolve slowly down the potential, so $w(z)$ starts from $-1$ 
   and increases later.  Almost all cases show this thawing feature as represented in Figs. 9 $\sim$ 12, however there are some cases which are a little bit different.  
     To the backward in time, the field $Q$ climbs up the potential, ceases and then falls down the potential 
     (typically the red curve of case e) in Figs. 9 (c) and 10 (a)).

  For the cosine type potential, there is the constraint that the following $Y$ must be negative.
  \begin{align}
Y=\frac{V''}{V} = -\frac{\cos\left(\frac{Q}{f}\right)}{f^2\left[\cos\left(\frac{Q}{f}\right)+1\right]}.
\label{PNGB_A}
 \end{align}

\begin{figure}[H]
\begin{minipage}{0.48\textwidth}
  \centering
   \includegraphics[clip,width=70mm]{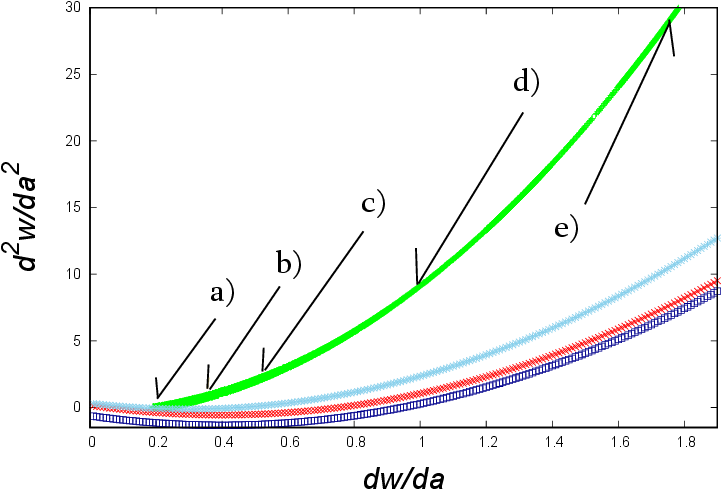}
  \subcaption{}
  \label{fig:9a}  
 \end{minipage}
 \hfill
 \begin{minipage}{0.48\textwidth}
  \centering
   \includegraphics[clip,width=76mm]{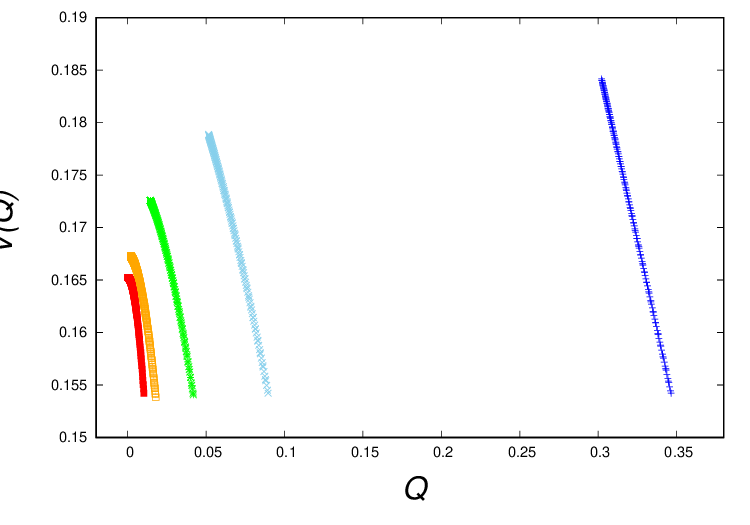}
  \subcaption{}
  \label{fig:9b}
 \end{minipage}
 \begin{minipage}{0.48\textwidth}
  \centering
   \includegraphics[clip,width=76mm]{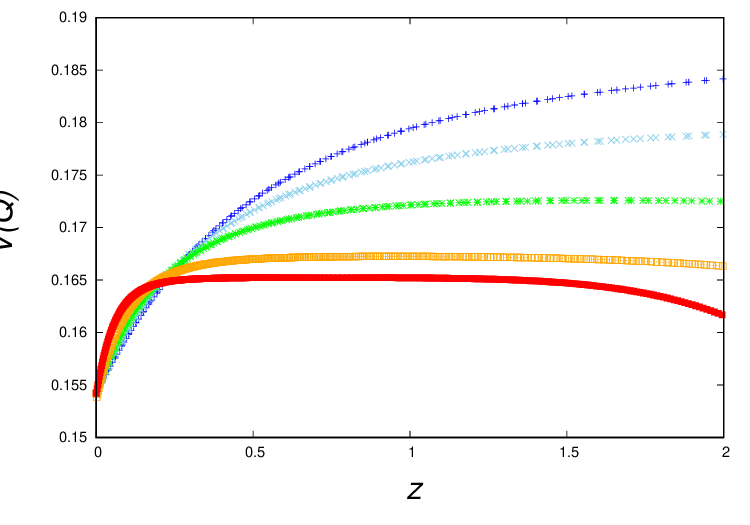}
  \subcaption{}
  \label{fig:9c}
 \end{minipage}
 \begin{minipage}{0.48\textwidth}
  \centering
   \includegraphics[clip,width=70mm]{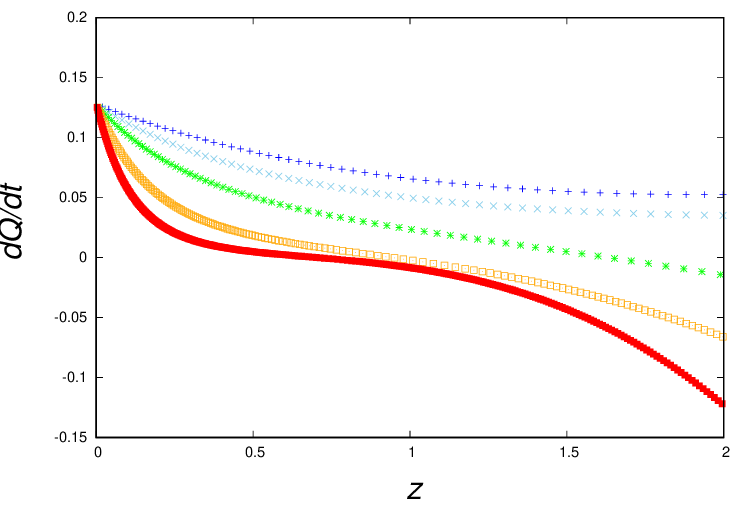}
  \subcaption{}
  \label{fig:9d}
 \end{minipage}
 \caption{Cosine potential.  (a) The same with Fig. 1 for the allowed green colored region for the cosine potential which must be noticed in different scales.
 The same with Fig. 1 for sky-blue, red and dark-blue curves.  The cases a) $\sim$ e) in Table IV are indicated. 
 (b) Downward movement of $Q$ from $z=2$ to $z=0$ in the cosine type potential $V(Q)$.  
 For each case, the parameter $f$ is not the same, so the form of the potential
  $V=M^4(\cos(Q/f)+1)$ is different. 
 However the value of potential $V$ at bottom is the same at present for every case as 
 $V_0=\rho_c\Omega_Q(1-\Delta/2)$ (Eq. (9)).  
 If the time normalization $t_0=(3/(4\pi \rho_c))^{1/2}$ is used, the value of $V$ at present is 
 $V_0=3\Omega_Q(1-\Delta/2)/(4\pi) \simeq 0.154$.   (c) Evolution of the potential height $V(Q)$ for each case is shown.  
 Three cases (c) d) and e)) have climbed at first ($z \sim 2$) and then slowed down.
 (d) Evolution of the velocity $dQ/dt$ for each case is presented.  Three cases (c) d) and e)) have started from $dQ/dt < 0$ at and then become positive.
   When $dQ/dt$ passes through $0$, $w$ takes the value of $-1$  which is noticed in Fig. 10 (a).  Especially case e) of red curve seems to be stagnating around some part (top?) of the potential. }   
\end{figure}

 
 \begin{table}[htb]
\centering
  \begin{tabular}{|l|r|r|r|r|r|} \hline
       & $dw/da$   &   $d^2w/da^2$  &  $f$       &  $Q_0$               & color \\ \hline \hline
    a) & 0.186  &  $-5.00\times10^{-2}$  &  0.223  &  0.346              & blue \\ \hline
    b) & 0.300  &  0.496 &  $9.74\times10^{-2}$  &  $8.97\times10^{-2}$  &  sky-blue \\ \hline
    c) & 0.500  &  2.00  &  $5.90\times10^{-2}$  &  $4.17\times10^{-2}$  &  green \\ \hline
    d) & 1.05   &  10.0  &  $3.10\times10^{-2}$  &  $1.77\times10^{-2}$  &  orange \\ \hline
    e) & 1.74   &  28.5  &  $1.98\times10^{-2}$  &  $1.04\times10^{-2}$  &  red \\ \hline 
  \end{tabular}
  \caption{The values of $dw/da$ and $d^2w/da^2$ for typical cases of the cosine potential in Figs. 9 and 10 (a).  
  The cases of a) to e)  are adopted along the green region in Figs. 1 and 9 (a).  
  It should be noted that the parameter $f$ and the value of $Q_0$ at present are different for each case.}  
\end{table}


\begin{figure}[ht]
\begin{minipage}{0.48\textwidth}
\centering
\includegraphics[clip,width=7cm]{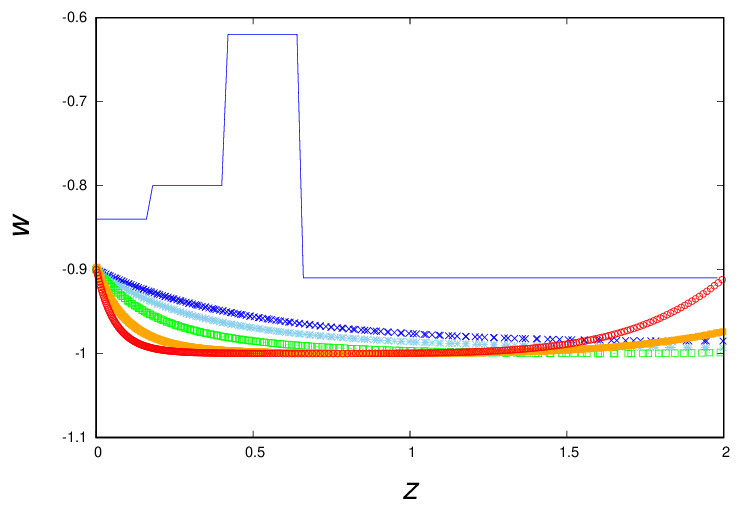}
\subcaption{}
 \end{minipage}
 \hfill
\begin{minipage}{0.48\textwidth}
\centering
\includegraphics[clip,width=7cm]{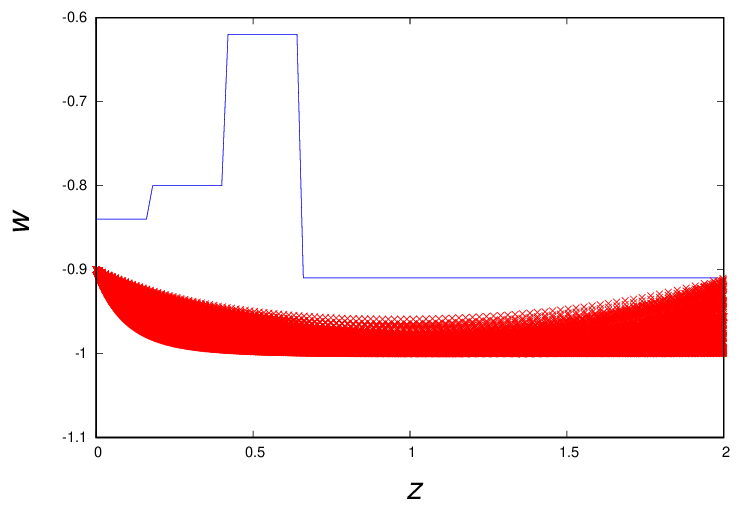}
\subcaption{}
\end{minipage}
\caption{(a) Under the observational constraint of blue lines, 
the allowed typical curves for the cosine type potential are displayed.    
It should be noted that the slope of curves is negative at $z=0$.  (b) Under the observational constraint, 
the allowed many cases for the cosine type potential in the green region in Figs. 1 and 9 (a) are displayed by red curves. 
It must be noted that, as the signature $d^2w/da^2$ is positive, the allowed curves are the downward convex (concave).}
\end{figure}


The above constraint reduced to the following equation 
   \begin{align}
Y =& -\frac{4\pi a^2}{3\Delta \Omega_Q}(1-\frac{\Delta}{2})^{-1}\left[\frac{d^2w_Q}{da^2}-\frac{1}{2\Delta} \left(\frac{dw_Q}{da}-\frac{3\Delta}{a}(1-2\Delta+\frac{(2-3\Delta)\Omega_Q}{6})\right)^2  \right. \nonumber \\
&+ \left.  \frac{3\Delta}{8a^2}\left(3((1-\Delta)(6+\Omega_Q)-\frac{\Omega_Q}{3})^2-16(7-6\Delta)(1-\Delta/2)\right)\right],  \label{YNEGATIVE}
\end{align}
which means that there is an allowed region, given by
\begin{align}
\frac{d^2w_Q}{da^2}&-\frac{1}{2\Delta} \left(\frac{dw_Q}{da}-\frac{3\Delta}{a}(1-2\Delta+\frac{(2-3\Delta)\Omega_Q}{6})\right)^2  \nonumber \\
&+  \frac{3\Delta}{8a^2}\left(3((1-\Delta)(6+\Omega_Q)-\frac{\Omega_Q}{3})^2-16(7-6\Delta)(1-\Delta/2)\right) > 0.  \label{YNEGATIVEII}
\end{align}

The critical boundary of the parabolic equation is shown by the sky-blue curve in Fig. 1.  
For this potential, the allowed parameter region in $dw_{Q}/da-d^2w_{Q}/da^2$ plane is shown by green in Figs. 1 and 9 (a).  
The typical cases a) $\sim$ e) with values of $dw/da, d^2w/da^2, f,$ and $Q_0$ are presented in Table IV.  
The example curves in Table IV are shown in Figs. 9 and 10.
The allowed region is positive in the vertical axis ($d^2w/da^2 > 0$) which are the curves of concave and/or downward convex.
The slopes of all lines are positive ($dw/da > 0$) which is understandable for the negative slope at $z=0$ in $dw/dz(=-(1+z)^{-2}dw/da < 0)$. 
These features are the same for the following Gaussian potential.

The scales must be noticed to be different in the vertical axes of Figs. 1, 2 (a) and  9 (a).  
Then the allowed region is large for this potential compared with the previous freezing type potential. 
 If we estimate the probability of the potential from the regional surface of $dw/da-d^2w/da^2$ space, the probability of this potential is large.

 \subsection{Gaussian type potential}

For the Gaussian type potential, it is interesting that the constraint $\sigma^2 > 0$ is the opposite region of $\alpha > 0$ 
  in $dw_Q/da -d^2w_Q/da^2$ plane of Fig. 1 \citep{5}.   For this potential, the allowed parameter region in $dw_{Q}/da-d^2w_{Q}/da^2$ plane is 
  presented by blue in Figs. 1 and 11 (a).  
  The typical cases a) $\sim$ e) with values of $dw/da, d^2w/da^2, \sigma,$ and $Q_0$ are presented in Table V.    
The example curves in Table V are shown in Figs. 11 and 12.
The allowed region in Fig. 11 (a) is positive in the vertical axis ($d^2w/da^2 > 0$) which are the curves of downward convex.
The slopes of all curves are positive ($dw/da > 0$) which is understandable for the negative slope at $z=0$ in $dw/dz(=-(1+z)^{-2}dw/da < 0)$.

One should notice that the scale height of $d^2w/da^2$ in Fig. 11 (a) is much higher than the similar figures.  If one estimate the probability 
of the potential from the allowed region surface in $dw/da-d^2w/da^2$ space, the allowed surface is the vastest compared with other potential regions.  
 The height of the allowed region becomes much higher than 4000 in numerical simulations.  The large value of  $d^2w/da^2$ means 
 the rapid increase of $w$ around $z \simeq 0$ (see Fig. 12 (a) especially red case e)).


\begin{figure}[H]
\begin{minipage}{0.48\textwidth}
  \centering
   \includegraphics[clip,width=70mm]{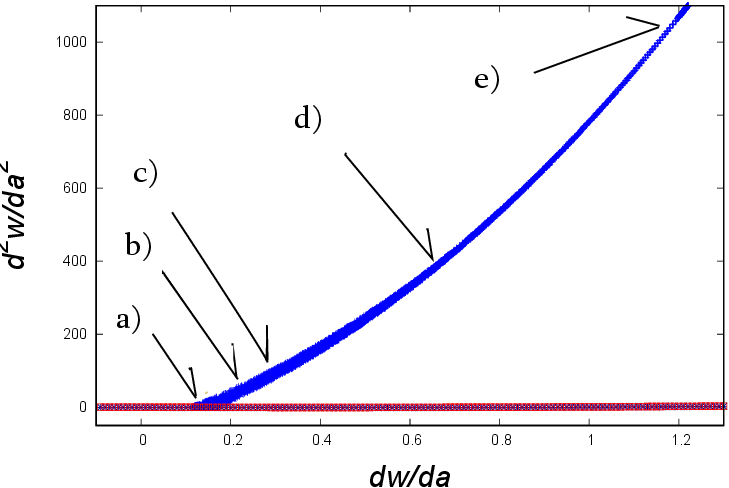}
  \subcaption{}
  \label{fig:11a}  
 \end{minipage}
 \hfill
 \begin{minipage}{0.48\textwidth}
  \centering
   \includegraphics[clip,width=70mm]{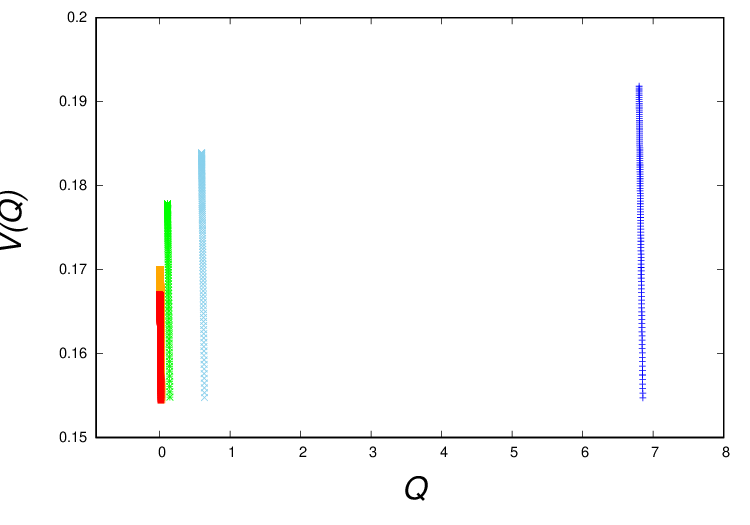}
  \subcaption{}
  \label{fig:11b}
 \end{minipage}
 \begin{minipage}{0.48\textwidth}
  \centering
   \includegraphics[clip,width=70mm]{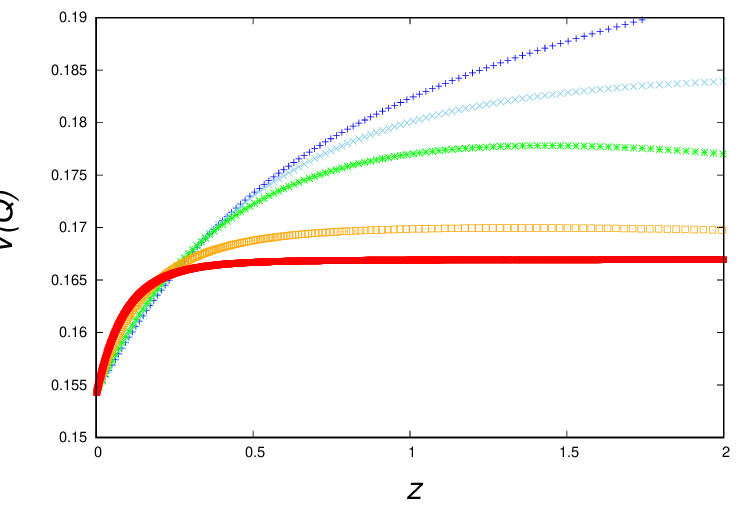}
  \subcaption{}
  \label{fig:11c}  
 \end{minipage}
 \hfill
 \begin{minipage}{0.48\textwidth}
  \centering
   \includegraphics[clip,width=70mm]{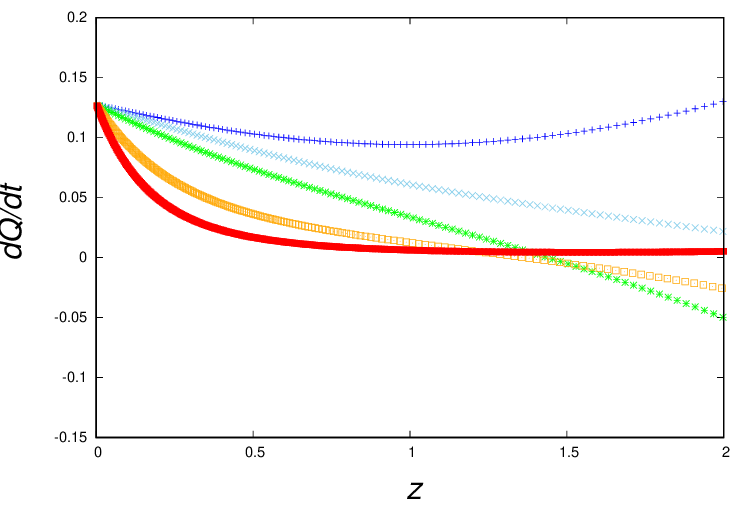}
  \subcaption{}
  \label{fig:11d}
 \end{minipage}
 
 \caption{Gaussian potential.  (a) The same with Fig. 1 for the allowed blue colored region 
 for the Gaussian potential which must be noticed for the different scale.
 The same with Fig. 1 for sky-blue, red and dark-blue curves, however they are overlapped.  The cases a) $\sim$ e) in Table V are indicated. 
 (b) Downward movement of $Q$ from $z=2$ to $z=0$ in the Gaussian potential.  
 For each case, the parameter $\sigma$ is not the same, so the form of the potential
  $V=M^4\exp(-Q^2/\sigma)$ is different. 
 However the value of potential $V$ at bottom is the same at present for every case as 
 $V_0=\rho_c\Omega_Q(1-\Delta/2)$ (Eq. (9)).  
 If the time normalization $t_0=(3/(4\pi \rho_c)^{1/2}$ is used, the value of $V$ at present is 
 $V_0=3\Omega_Q(1-\Delta/2)/(4\pi) \simeq 0.154$.   (c) Evolution of the potential height $V(Q)$ for each case is shown.  
 Two cases (c) and d) have climbed at first ($z \sim 2$) and then slowed down.
 (d) Evolution of the velocity $dQ/dt$ for each case is presented.  Two cases c) and d) have started from $ dQ/dt < 0$ at $z=2$ and then become positive.  
 When $dQ/dt$ passes through $0$, $w$ takes the value of $-1$  which is noticed in Fig. 12 (a). 
 Case e) seems to be stagnating around some part (top?) of the potential. }   
\end{figure}

 
 \begin{table}[htb]
\centering
  \begin{tabular}{|l|r|r|r|r|r|} \hline
       & $dw/da$      &      $d^2w/da^2$ &      $\sigma$ & $Q_0$   & color \\ \hline \hline
    a) & 0.118 &  0.700 & 1.85 & 6.85   & blue \\ \hline
    b) & 0.158 &  10.0 & 0.543 & 0.637  & sky-blue \\ \hline
    c) & 0.241 &  50.0 & 0.247 & 0.145  & green \\ \hline
    d) & 0.673  & 400.0 & $8.78\times10^{-2}$ & $2.80\times10^{-2}$     & orange \\ \hline
    e) & 1.12  &  950 & $5.70\times10^{-2}$ & $1.61\times10^{-2}$     & red \\ \hline 
  \end{tabular}
  \caption{The values of $dw/da$ and $d^2w/da^2$ for typical cases of the Gaussian potential in Figs. 11 and 12 (a).  
  The cases of a) to e)  are adopted along the blue region in Fig. 11 (a).  
  It should be noted that the parameter $\sigma$ and the value of $Q_0$ at present are different for each case.}  
\end{table}


\begin{figure}[ht]
\begin{minipage}{0.48\textwidth}
\centering
\includegraphics[clip,width=7cm]{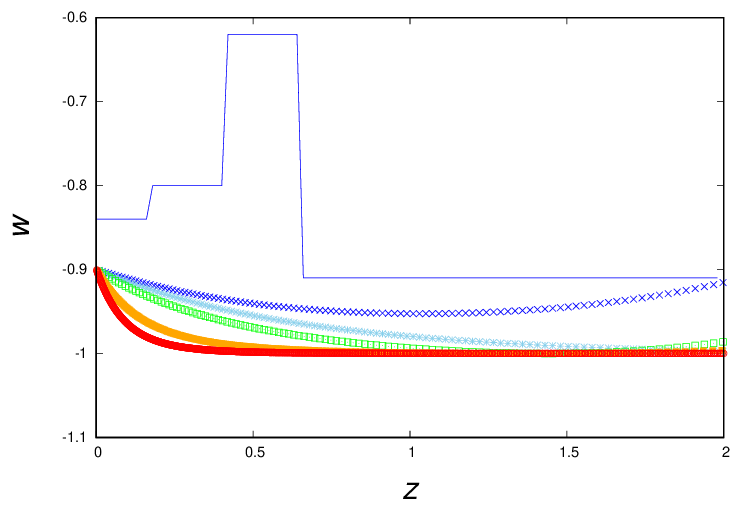}
\subcaption{}
\end{minipage}
\hfill
\begin{minipage}{0.48\textwidth}
\centering
\includegraphics[clip,width=7cm]{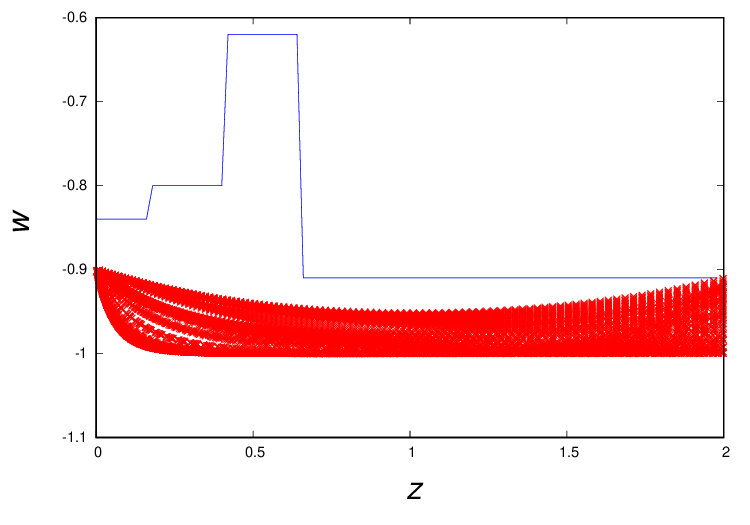}
\subcaption{}
\end{minipage}
\caption{(a) The constraint from the observation is shown by blue lines.  Under these constraint, 
the allowed typical cases in Table V for the Gaussian type potential are displayed. (b) Under the observational constraint, 
the allowed almost all cases for the Gaussian type potential in the blue region in Figs. 1 and 11 (a) are displayed by red curves. } 
\end{figure}

Although it is difficult to deny the possibility of the large value of $d^2w/da^2$, 
 it seems to be low for its rapid increase of $w$ around $z \simeq 0$ (see Fig. 12 (a) especially red case e)). 
  At least such a rapid change of $w$ will be checked in near future.

\section{Dependence on the parameter $\Delta$}


\begin{figure}[H]
\begin{minipage}{0.48\textwidth}
\centering
\includegraphics[clip,width=7cm]{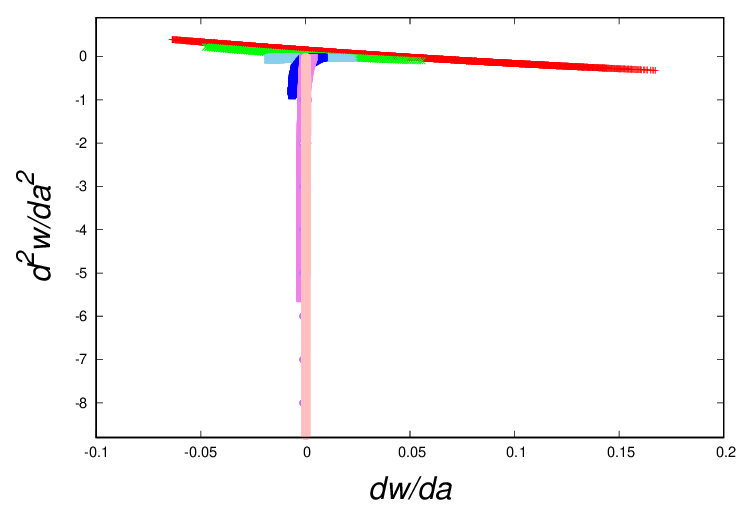}
\subcaption{}
\end{minipage}
\hfill
\begin{minipage}{0.48\textwidth}
\centering
\includegraphics[clip,width=7cm]{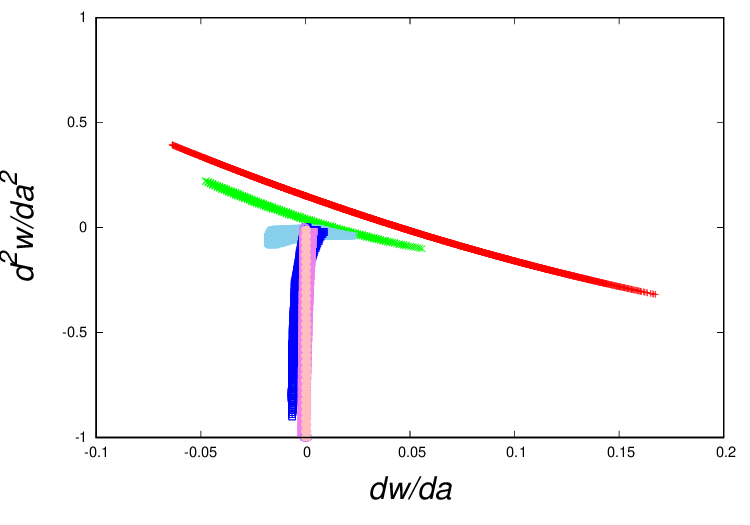}
\subcaption{}
\end{minipage}
\caption{Inverse power law potential.  (a) The same with Fig. 1 for the inverse power law potential except for the value of $\Delta$ and the red region is for $\Delta$=0.1.  
For $\Delta$=0.03, 0.01, and 0.003, the regions are colored by green, sky-blue, and blue respectively.  For $\Delta$=0.001, 0.0003, and 0.0001, the regions are colored 
by violet, purple, and pink, respectively.  The cases around the vertical pink region are investigated in Fig. 14 (a) and (b).  (b) The same with Fig. 13 (a) for 
the inverse power law potential except for the scale in the vertical axis.  
}  
\end{figure}

 \hspace{5mm}The allowed regions in Fig. 1 depend on the parameter $\Delta(=w+1)$.  
 Then we investigate the allowed regions by changing the values of $\Delta$ from 0.1, to  0.0001, 
 The results for $\Delta$=0.1, 0.03, 0.01, 0.003, 0.001, 0.0003, and 0.0001 are represented in Figs. 13 and 15-18
 by red, green, sky-blue, blue, violet, purple, and pink, respectively.

For the inverse power potential, the results are presented in Fig. 13  (a).  The enlarged allowed region is presented in Fig. 13 (b).  
The two cases are presented in Figs. 14 (a) and (b).  
The evolution $w$ with time $z$  is shown in Fig. 14 which is the same with Fig. 4 except for the value of $\Delta=0.0001$.

To analyze the pink region, we have presented two cases in Fig. 14 (a) and (b).  
The red curve is within an allowed region (pink) in Fig. 13  and green curve is out of the allowed region (right-hand side).
The value of $d^2w/da^2$ is  about $-10$ for both cases.

\begin{figure}[t]
\begin{minipage}{0.48\textwidth}
\centering
\includegraphics[clip,width=7cm]{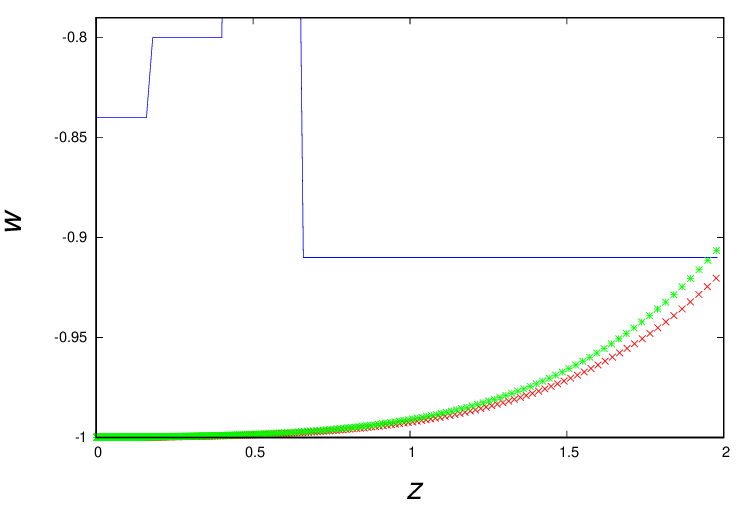}
\subcaption{}
\end{minipage}
\hfill
\begin{minipage}{0.48\textwidth}
\centering
\includegraphics[clip,width=7.6cm]{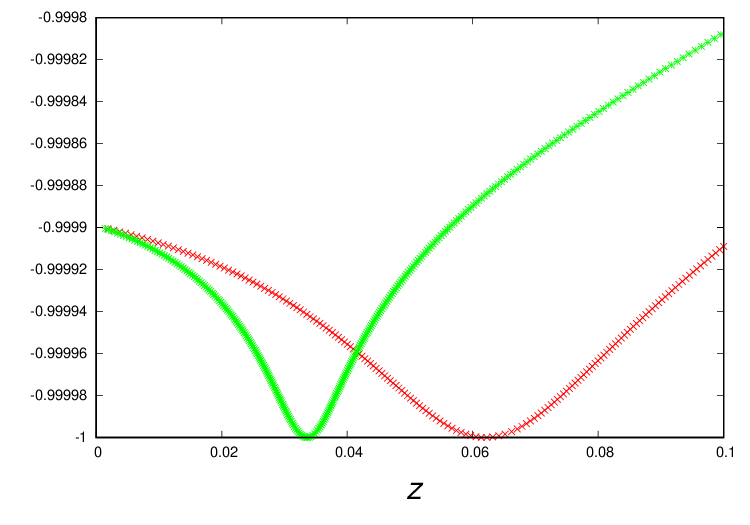}
\subcaption{}
   
\end{minipage}
\caption{(a) The same with Fig. 4 (a) for the inverse power law potential except for the value of $\Delta=0.0001$,  
The red curve is within an allowed region (pink) in Fig. 13 and green curve is out of allowed region (right-hand side).
The value of $d^2w/da^2$ is about $-10$ for both curves.  The $w$ value of green one at $z$=2 is greater than $-0.91$ which is not allowed.  
The left-hand side of the pink region is forbidden by the constraint $adw/da+6\Delta(1-\Delta/2) \ge 0$
 which could be understood  from Eq. (12).   (b) The same with Fig. 14 (a), except for the horizontal range where the range $ 0 \le z \le 0.1$ is enlarged.  Both curves show the similar features.
Toward the past, the field $Q$ decreases which means that the field $Q$ climbs up the potential $V=M^{4+\alpha}/Q^{\alpha}$ 
and then returns back down the potential. 
At the turnover the value of $dQ/dt$ becomes null and $w$ becomes $-1$.  }
\end{figure}

In Fig. 14 (a), the $w$ value of green one at $z=2$ is greater than $-0.91$ which is not allowed by the observation (blue line).  
The detailed features around $z \simeq 0.05$ is shown in Fig. 14 (b).  In Fig. 14 (b), it is almost the same with Fig 14 (a), 
except for the horizontal range where the range $ 0 \le z \le 0.1$ is enlarged.  
Both curves show the similar features.

\begin{figure}[h]
\centering
\includegraphics[clip,width=10cm,height=6cm]{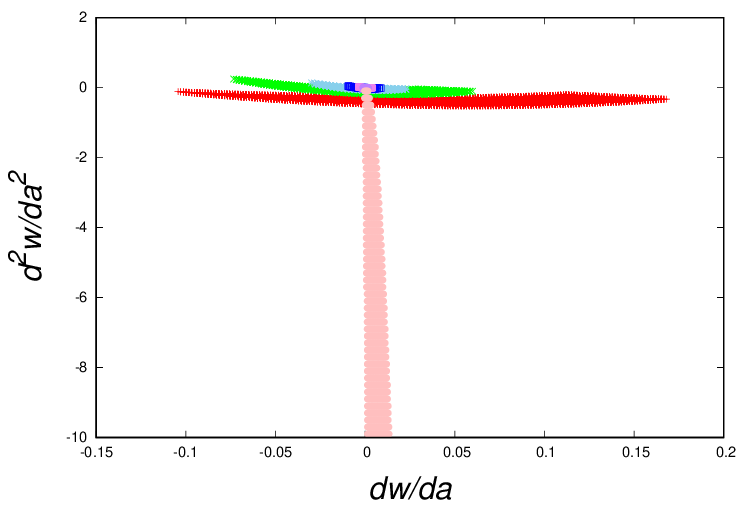}
\caption{The same with Fig. 1 for the exponential type potential except for the value of $\Delta$.  The red region is for $\Delta$=0.1.  
For $\Delta$=0.03, 0.01, and 0.003, the regions are colored by green, sky-blue, and blue respectively.
For $\Delta$=0.001, 0.0003, and 0.0001, the regions are colored by violet, purple, and pink, respectively. }  
\end{figure}

\begin{figure}[b]
\centering
\includegraphics[clip,width=10cm,height=6cm]{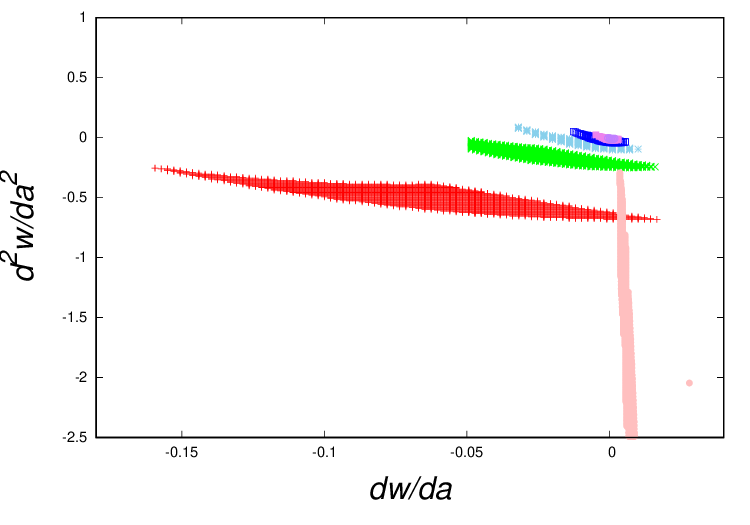}
\caption{The same with Fig. 1 for the mixed type potential except for the value of $\Delta$.  The red region is for $\Delta$=0.1.  
For $\Delta$=0.03, 0.01, and 0.003, the regions are colored by green, sky-blue, and blue respectively. 
For $\Delta$=0.001, 0.0003, and 0.0001, the regions are colored by violet, purple, and pink, respectively.  }  
\end{figure}

Toward the past, the field $Q$ decreases (not shown here) which means that the field $Q$ climbs up the potential $V=M^{4+\alpha}/Q^{\alpha}$ 
and then returns back down the potential. 
At the turnover the value of $dQ/dt$ becomes null and $w$ becomes $-1$.  
Then the field $Q$ comes down the potential which shows the bouncing feature of $w$ with time ($z$).

If the second derivative of $w$ with $z$ is negative, it is the curve of convex upward which is observed far from the bouncing region.  
Around the bouncing region, it is the curve of convex downward (concave) where the second derivative of $w$ with $z$ is positive, 
even though the second derivative $w$ with $a$ is negative at $z=0$ 
($\frac{d^2w}{dz^2}=\frac{2}{(1+z)^3}\frac{dw}{da}+\frac{1}{(1+z)^4}\frac{d^2w}{da^2} \ge  0$ ).

The left-hand side of the allowed pink region is forbidden by the constraint $adw/da+6\Delta(1-\Delta/2) \ge 0$ which could be understood in  Eq. (12).

For the exponential potential, the results are presented in Fig. 15. 
  For the mix type potential, the results are presented in Fig. 16.  These freezing type potentials have almost a similar feature that
   the pink region extends downward ($d^2w/da^2 \ll -1$).  As explained in Figs. 14 (a) and (b), 
   the negative value of $d^2w/da^2 $ means that the field $Q$ climbed up the potential (backward to time), stopped ($w=-1$), 
   and fall down the potential.  The extreme negative value of  $d^2w/da^2 $ means 
   the bouncing point ($z \simeq 0.06$ in Fig. 14 (b) ) approaches to $z \simeq 0$, then the extreme convex feature appeared in 
   the redshift evolution of $w(z)$.

\begin{figure}[H]
\begin{minipage}{0.48\textwidth}
\centering
\includegraphics[clip,width=7cm]{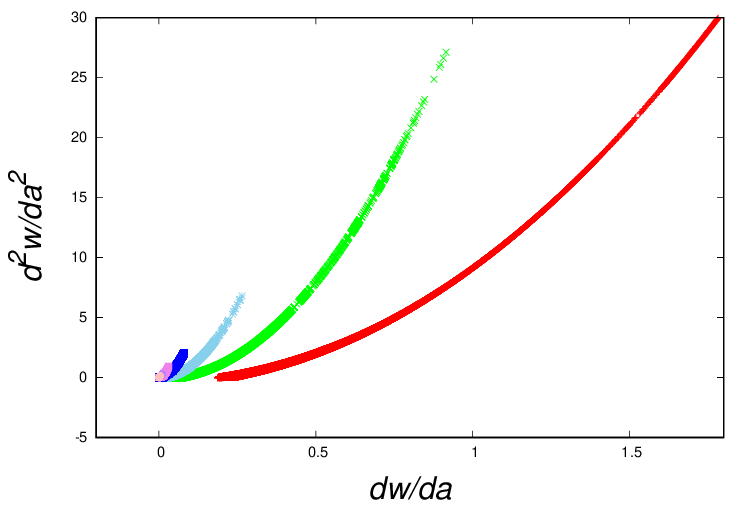}
\subcaption{}
\end{minipage}
\hfill
\begin{minipage}{.48\textwidth}
\centering
\includegraphics[clip,width=7cm]{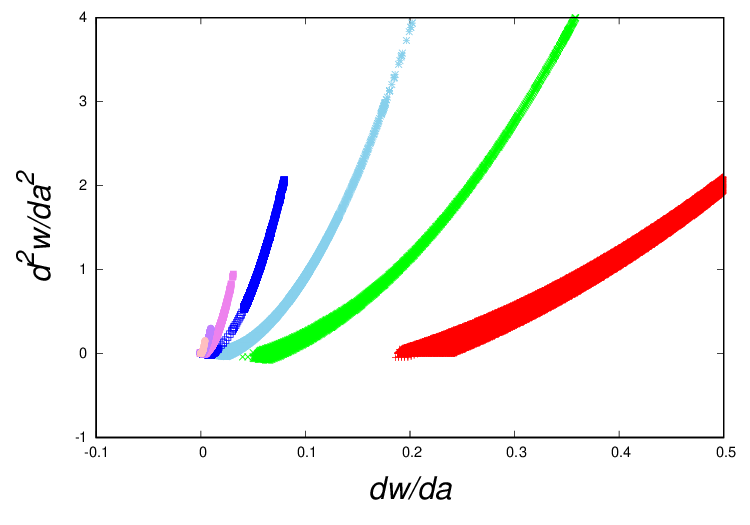}
\subcaption{}
\end{minipage}
\caption{(a) The same with Fig. 1 for the cosine type potential except for the value of $\Delta$.  The red region is for $\Delta$=0.1.  
For $\Delta$=0.03, 0.01, and 0.003, the regions are colored by green, sky-blue, and blue respectively. 
For $\Delta$=0.001, 0.0003, and 0.0001, the regions are colored by violet, purple, and pink, respectively.  
(b) The same with (a) except for the region around (0, 0) enlarged.  The region surrounded 
by red curve and blue or purple curves has the possibility for allowed case, if one take the appropriate value of $\Delta$ or continuous value of 
$\Delta (0.0001 \leq \Delta \leq 0.1)$.}    
\end{figure}

For the cosine potential, the results are presented in Fig. 17.  The region surrounded 
by red curve and blue or purple curves has the possibility for allowed case if one takes the appropriate value of $\Delta$ or continuous value of 
$\Delta (0.0001 \leq \Delta \leq 0.1)$.

For the Gaussian type potential, the results are presented in Fig. 18.  If one estimate the probability of 
the potential from the surface of $dw/da-d^2w/da^2$ space, the Gaussian potential has the largest possibility to represent the quintessence field,
 because the region bounded by the red and pink curves is very vast compared with other potentials.  
The second large possibility is the cosine type potential bounded by red and pink curves in Fig. 17.

In the limit $\Delta \rightarrow 0$ the allowed region for every potential includes the part of $dw/da \simeq 0$ and $d^2w/da^2 \simeq 0$.  
In this limit, it is hard  to discern the potential type in the quintessence scenario as well as the cosmological constant model ($w=-1$).

\begin{figure}
\begin{minipage}{.48\textwidth}
\centering
\includegraphics[clip,width=7cm]{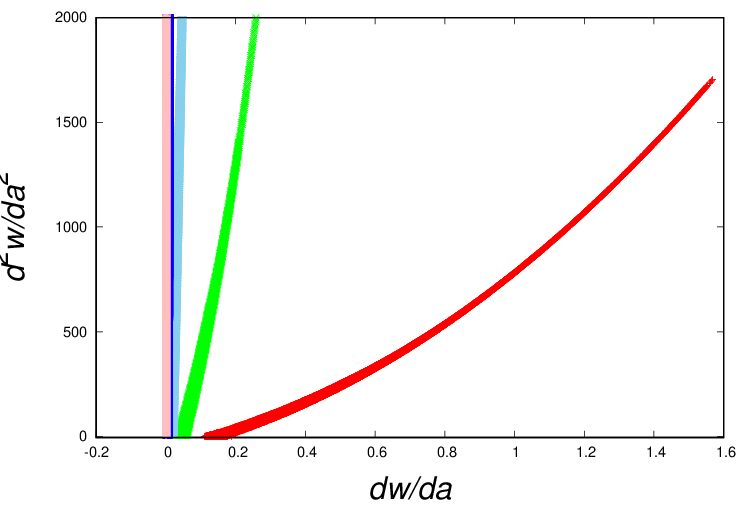}
\subcaption{}
\end{minipage}
\hfill
\begin{minipage}{.48\textwidth}
\centering
\includegraphics[clip,width=7cm]{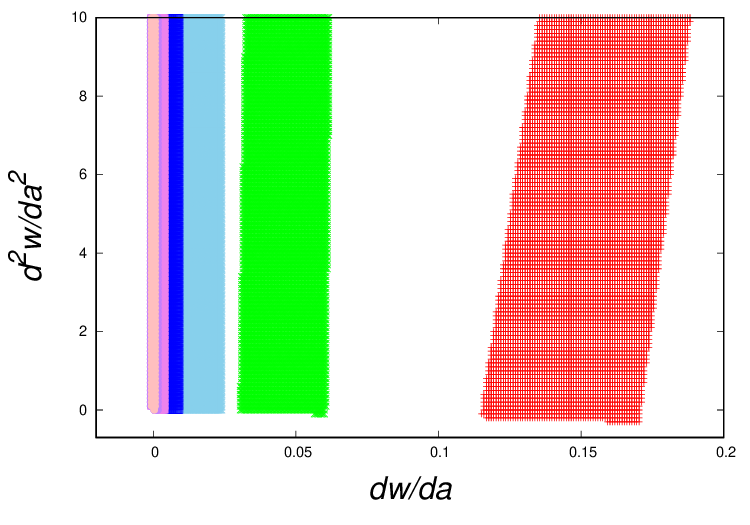}
\subcaption{}
\end{minipage}
\caption{(a) The same with Fig. 1 for the Gaussian type potential except for the value of $\Delta$.  The red region is for $\Delta$=0.1.  
For $\Delta$=0.03, 0.01, and 0.003, the regions are colored by green, sky-blue, and blue, respectively. 
For $\Delta$=0.001, 0.0003, and 0.0001, the regions are colored by violet, purple, and pink, respectively.  (b) The same with Fig. 26, except for the small value part of $dw/da$ and $d^2w/da^2$ are enlarged.  
The regions covered by violet, purple and pink are almost overlapped,  }  
\end{figure}

\section{Another Observational Constraint}
\hspace{5mm} We take another observational constraint in the paper (Fig. 5; Planck+WL+BAO/RSD) \citep{20} which extends until $z=5$ for $w$ shown in Fig. 20 by the green curve.  
The constraint in Fig. 20 by blue lines ($ z \le 2$) is the one in the paper (Fig. 7: Planck+BSH) \citep{20}.  
To some extent, both constraints are similar around $z \simeq 0.6$, 
however the adopted constraint of green curve in this section has 
the long constraint until $z=5$.
  The allowed regions for the investigated potentials shrink as presented in Fig. 19 where the allowed regions are much smaller than those in Fig. 1.
The example curves for the case of the inverse power potential are shown in Fig. 20 by red curves.  
The constraint for the large redshift ($z \le 5$) has the effect 
  on the decrease of the allowed region for each potential in $dw/da-d^2w/da^2$ space. 
  It is noticed that 
the constraint of $z \simeq 5 $ has the effect to shrink the allowed region, 
as the brown colored region in Fig. 1 has decreased a little bit in Fig. 19.

\begin{figure}[H]
\centering
\includegraphics[clip,width=10cm,height=7cm]{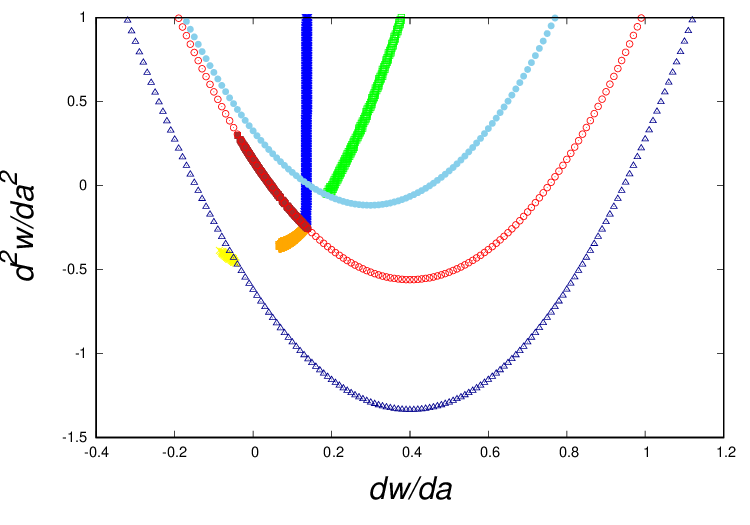}
\caption{The allowed regions for the adopted potentials under another observational constraint which is the one in the paper (Fig. 5) \citep{20}.
Compare with Fig. 1 where the observational constraint is the one in the paper (Fig. 7) \citep{20}.  The parabolic curves are the same with Fig. 1.  Each shrunk colored region corresponds to the same potential as Fig. 1, respectively.  
The brown curve-like region is for the inverse power potential.  
 The orange region is for the exponential potential.  
The yellow colored region shows the allowed region for the mixed type potential.  
The blue region is for the Gaussian potential and the green region is the allowed region for the cosine potential.
To some extent, both constraints are similar, however the adopted in this section has the long constraint until $z=5$ 
and the allowed regions have shrunk. In the simulation, we take $\Delta=0.1$ and $\Omega_Q=0.68$. }   
\end{figure}

\begin{figure}{H}
\centering
 \includegraphics[clip,width=10cm,height=7cm]{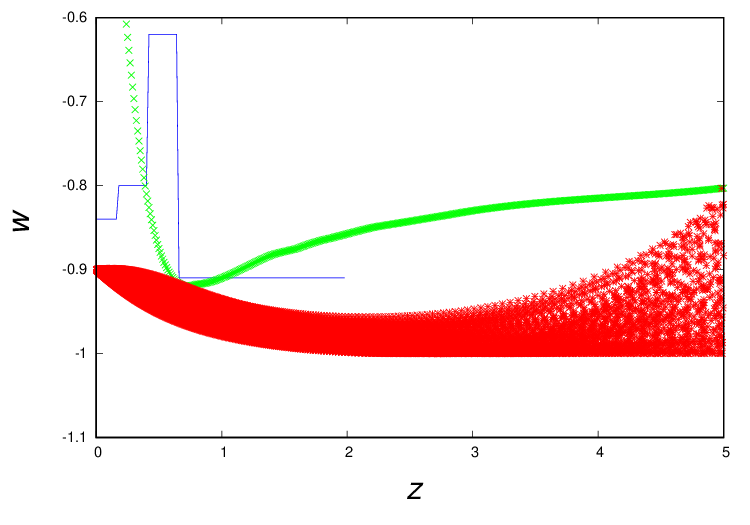}
\caption{The same with Fig. 5, however the new constraint until $z=5$ is shown by green curve where the previous constraint is shown by blue lines.  
The evolutions of $w$ for many cases of the inverse power potential are shown by red color curves. It is noticed that 
the constraint of $z \le 5 $ has the effect to shrink the allowed region, as the brown colored region in Fig. 1 has decreased a little bit in Fig. 19.}   
\end{figure}

It is apparent that other colored region in Fig. 1 has decreased in Fig. 19.  
On the other hand, the allowed region exists for any potential, even though the region has shrunk.

\section{Results and Discussion}
\hspace{5mm}We study the accelerating universe in terms of the time evolution of the equation of state $w(z)$ 
for the thawing and freezing potentials in the quintessence scenario.

The investigated exact potentials have two parameters for their forms. To determine the parameters from the observation, we need $ dw/da$ and $d^2w/da^2$ and
 we could not find out the papers to refer this point except \citep{5, 6}. 
If we know the exact potential with two parameters, it becomes possible to simulate the evolution of the scalar field with 
the two boundary conditions $Q$ and $dQ/da$ (it is equivalent to have $w_0$ and $\Omega_Q$ (Eqs. (9) and (10)), 
being the second differential equation, which determines the evolution of $Q$. 
In the end, it is necessary to observe four values $w_0, \Omega_Q, dw/da_0$, and $d^2w/da^2$, which determine the potential parameters and evolution of the scalar field. 

The main procedure of this paper is the following: By adopting the values of 
the derivatives $dw/da$ and $d^2w/da^2$, the characteristic parameters of each potential can be fixed 
and then the theoretical time evolution of the scalar field $Q$ in the potential is numerically calculated from the present to the past 
which is time reversible. 
The time evolutions of $w(z)$ which pass through the observational constraint are selected and those values of $dw/da$ and $d^2w/da^2$ are 
presented as an allowed region for each scalar potential which are summarized in Fig. 1.\\



By comparing the observations of $w(z)$ variation with the simulations, we can discriminate the quintessence potentials through Fig. 1.  
For the moment as there is a lot of uncertainty in the observations, every investigated potential has the possibility to explain them, 
however the form parameters of each potential have been limited which are related to $dw/da$ and $d^2w/da^2$ as stated in the paper and especially Appendix B.
The parameter space of $dw/da-d^2w/da^2$ are limited as shown in Fig. 1. 
If future observations become improved and more precise in $dw/da - d^2w/da^2$ space, it will be possible to discriminate the potential types and potential parameters.\\


 There are some cases which show the thawing feature in the freezing type potentials.  
It has been assumed that the scalar field flows down the freezing type potential.  However, under the observational constraint of $z ( \leq 2)$, 
there is a possibility that scalar field climbs up the potential and stop there, where $w$ becomes $-1$, and then flows down 
where $w$ increases from $ -1$.
  If we consider that such initial conditions are not natural, there will be no allowed $dw/dz-d^2w/da^2$ space for the freezing potentials. 
   It is due to the observational constraint that $w$ must be lower $ -0.91$ around $ 0.65 \leq z \leq 2$ and $w$ is $-0.9$ at present ($z=0$).  
  If we accept the rather unnatural conditions that $Q$ climbs up the potential, situated there and roles down for the freezing potentials, 
  there is a possibility that freezing type potentials could survive.

  Even for thawing potential, there are rather unnatural solutions that $Q$ climbs up the potential, situated there and roles down.
  These solutions are investigated by Swaney and Scherrer \citep{19}.
  We accept those solutions from the phenomenological point that the observations do not exclude such solutions for the moment.  
  It will be a future problem why such initial conditions are realized if such solutions are found in observations. \\

If one estimate the probability of 
the potential from the surface region of $dw/da-d^2w/da^2$ space, the Gaussian potential is the first, because the region bounded by the red 
and pink curves  in Fig. 18 (a) is the largest compared with other potentials.  
The second possibility is the cosine type potential bounded by red and pink curves in Fig. 17. 
Under the observations \citep{20},  there are two ample space regions in $dw_Q/da-d^2w_Q/da^2$ space for thawing type potentials 
(Gaussian type and cosine type). 
Only the relatively small regions are allowed for freezing type potentials.

Although there is a lot of uncertainty at the moment, it seems to be preferable for the thawing model against the freezing model 
under the comparison with the numerical results and the observations.  It should be stressed that the detailed observation of $dw/da$, and $d^2w/da^2$ will determine 
the potential type of the quintessence scenario as shown in Figs. 1, 9 (a), 11 (a), and 19, respectively.

About the other observational constraint ($z \le 5$) in $\S 5$, we hope the constraint becomes much severe and extends to higher $z$, 
however the observational constraint must be very hard where the ratio of dark energy to matter is smaller than $10^{-2}\sim 1\%$ at $z=5$
 as $ \Omega_Q(z)/\Omega_m(z) \simeq \Omega_Q(0)/(\Omega_m(0)\times (1+z)^3 )$.   \\

There is a lot of many works related dark energy under a quintessence scenario and this work concentrates on this model. 
Although one parameter ($\Delta$) formula is investigated in the reference \citep{16} and some interesting features are predicted, 
our work introduces two more parameters and various other possibilities are predicted under the observational constraint \citep{20}.
There are many other trials to investigate $w(z)$ of dark energy and they have the original perspective in the quintessence scenario 
\citep{10,11,12,13,14,15,16,17,18,19,20,21,22,23}. 
 We believe that it is urgent to determine $\Delta > 0$ for quintessence and other models.  Moreover, if $ \Delta  \leq  0 \ (w \leq -1),$ 
we have to consider utterly different scenarios, \citep{1,24,25,26,27,28}.   

We  hope that it will be improved in detail the observation of $w(z)$ variation as presented by Planck (2016) \citep{20} in the following projects; 
BOSS: Baryon Oscillation Spectroscopic Survey,
DES: Dark Energy Survey,
WFIRST: Wide-Field Infrared Survey Telescope, Euclid (ESA satellite),
and CMB (COrE) (ESA satellite) projects \citep{28}.\\

\vspace{1cm}

{\Large{\bf{Acknowledgements}}} \\

\hspace{5mm} One of the authors (T. H) would like to thank Dr. Yutaka Itoh and  
Dr. Kei Nishi for the discussion about the problems in this paper.  \\

\vspace{0.2cm}

{\Large{\bf{Appendix  A} : Taylor Expansion Approximation} } \\

As we have calculated the third derivative for each potential \citep{5}, 
 we could estimate the third derivative ($d^3w/da^3$) from the observed first derivative ($dw/da$) and second derivative ($d^2w/da^2$).
  As we have expanded $w(a)$ in Taylor expansion \citep{20} given by
\begin{align}
w(a)=w_0+w_a(1-a)+\frac{1}{2}w_{a2}(1-a)^2 +\frac{1}{3!}w_{a3}(1-a)^3 , \label{act1}
\end{align}
where $a, w_0,  w_a, w_{a2}$ and $ w_{a3} $ are the scale factor ($a=1$ at current), the current value of $ w(a) $, the first, second and third derivatives of $ w(a)$ as $w_a=-dw/da$   
$w_{a2}=d^2 w/da^2$ and $w_{a3}=-d^3 w/da^3$, respectively.
We take it as the following relation of $z=(1-a)/a$, being $1+z=a_0/a$ and $a_0=1$.
\begin{align}
w(z)=w_0-w_0(z/(1+z))+\frac{1}{2}w_{a2}(z/(1+z))^2 -\frac{1}{3!}w_{a3}(z/(1+z))^3 , \label{act1a}
\end{align}
where we neglect the higher expansion term.  It is different from the treatment in the text 
where the exact evolution of $w$ under each potential is calculated. \\

\begin{figure}[ht]
\includegraphics[clip,width=10cm,height=6cm]{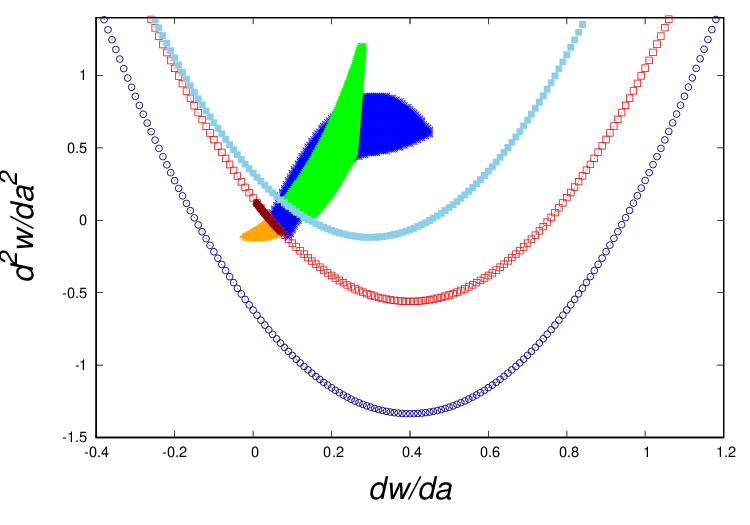}
\caption{The same with Fig. 1, however the evolution $w$ is calculated by Taylor expansion with the first 4 terms 
(including until the third derivative). The allowed regions for the inverse power law, exponential, cosine, and Gaussian potentials are 
colored by brown, orange, blue and green, respectively.  There is no region for the mixed type potential.  
In the simulation, we take $\Delta=0.1$ and $\Omega_Q=0.68$. }   
\end{figure}

The results are presented in Fig. 21 which is analogous in Fig. 1.
In Fig. 21, the allowed regions for inverse power, exponential, cosine, and Gaussian potentials are displayed 
by brown, orange, blue and green color, respectively.  There is no allowed region for the mixed type potential in this Taylor expansion formalism. 
In the text, the evolution of $Q$ is exactly calculated.  
Although Taylor expansion is the approximation, however there are some similarities between Fig.21 and Fig. 1, such as   
the allowed region for the inverse power law potential is almost on the one-dimensional curve. \\   
\vspace{0.1cm}

{\Large{\bf{Appendix  B} : Derivation of parameters}} \\

In this appendix, it is outlined the procedure for the derivation of parameters for each potential.
The equation number Eq. (x) in the reference (Muromachi {\it et al.} \citep{5}) is referred as Eq. (M x) in this appendix.\\

{\bf{B. 1}   Exponential Potential} 

  By adopting $dw/da$ in Eq. (12) and/or (M 38) , $V'/V=-\beta M/Q^2$ is estimated.  Using this and $d^2w/da^2$, $Q$ is derived in Eq. (M 42). 
So $\beta M/Q $ is taken from Eq. (M 38), $M$ is written in Eq. (M 43).   Then $\beta$ is obtained from Eq. (M 35).\\

{\bf{B. 2 }  Mixed type Potential}

The parameter $\zeta$ in the reference (\citep{5}) is taken $\zeta=4\pi$ in this paper. Originally $\zeta$ is the constant $(\zeta=4\pi)$ \citep{9}.  

  By adopting $dw/da$, $V'/V=-\gamma/Q +8\pi Q=X$ is estimated as in Eq. (M 49) and (M 50).   Taking this and $d^2w/da^2$, $Y=V''/V$ is derived.  
Using the relation $Y=X^2+\gamma/Q^2+8\pi$, $\gamma/Q^2$ is obtained as in Eq. (M 53).  Dividing Eq. (M 50) by $Q$, $X/Q$ is estimated. 
As we know $X$, $Q$ is derived.  Then it is easy to get $\gamma$, through $X$ or $Y$.\\
  
{\bf{B. 3}   Cosine Potential}

 By adopting $dw/da$ in Eq. (12) and/or (M 60), $V'/V=-f^{-1}\tan(Q/(2f))=X$ is estimated.  
 Taking this and $d^2w/da^2$, $Y=V''/V$ is derived in Eq. (M 63).   Through $X$ and $Y$, $f=(X^2-2Y)^{-1/2}$ is obtained.  
 Then $Q$ is easily taken in Eq. (M 65).  $M$ is obtained through the relation $V=\rho_c\Omega_Q(1-\Delta/2)=M^4(\cos(Q/f)+1)$.  \\
 
{\bf{B. 4}   Gaussian Potential}
 
 By adopting $dw/da$, $V'/V=-2Q/\sigma^2$ is estimated as in Eq. (M 70) and (M 71).  By taking this and $d^2w/da^2$, $\sigma^2$ is obtained in Eq. (M 74).
 Then $Q$ and $M$ are specified by $V'/V$ and $V=\rho_c\Omega_Q(1-\Delta/2)=M^4\exp(-Q^2/\sigma^2)$ as in Eq. (9).

\end{document}